\documentclass[structabstract,longauth]{aa}

\usepackage{graphicx}
\usepackage{txfonts, longtable, dcolumn, ctable}
\usepackage[T1]{fontenc}
\usepackage{natbib}
\bibpunct{(}{)}{;}{a}{}{,}

\begin{document}

   \title{Asteroids' physical models from combined dense and sparse photometry and scaling of the YORP effect by the observed obliquity distribution}

   \author{J. Hanu{\v s}
	  \inst{1*}
           \and
          J. {\v D}urech\inst{1}
           \and
          M. Bro{\v z}\inst{1}
           \and
	  A.~Marciniak\inst{2}
           \and
          B.~D.~Warner\inst{3}
           \and
          F.~Pilcher\inst{4}
           \and
	  R.~Stephens\inst{5}
           \and
	  R.~Behrend\inst{6}
	   \and
	  B.~Carry\inst{7}
	   \and
	  D.~\v Capek\inst{8}
% ...
   }

   \institute{
	     Astronomical Institute, Faculty of Mathematics and Physics, Charles University in Prague, V~Hole{\v s}ovi{\v c}k{\'a}ch 2, 18000 Prague, Czech Republic\\
	     $^*$\email{hanus.home@gmail.com}
         \and
	     Astronomical Observatory, Adam Mickiewicz University, S\l oneczna 36, 60-286 Pozna\' n, Poland 
         \and
             Palmer Divide Observatory, 17995 Bakers Farm Rd., Colorado Springs, CO 80908, USA 
	 \and
	     4438 Organ Mesa Loop, Las Cruces, NM 88011, USA 
	 \and
	     Goat Mountain Astronomical Research Station, 11355 Mount Johnson Court, Rancho Cucamonga, CA 91737, USA 
	 \and    
	     Geneva Observatory, CH-1290 Sauverny, Switzerland 
	 \and
	     European Space Astronomy Centre, Spain, P.O. Box 78, 28691 Villanueva de la Ca{\~ n}ada, Madrid, Spain 
	 \and
	     Astronomical Institute of the Academy of Sciences, Fri\v cova 298, 25165 Ond\v rejov, Czech Republic 
% ...
\and
	     ANS Collaboration, c/o Osservatorio Astronomico di Padova, Sede di Asiago, 36032 Asiago (VI), Italy 
% ...
}

   \date{Received x-x-2012 / Accepted x-x-2012}
 
  \abstract
   {The larger number of models of asteroid shapes and their rotational states derived by the lightcurve inversion give us better insight into both the nature of individual objects and the whole asteroid population. With a larger statistical sample we can study the physical properties of asteroid populations, such as main-belt asteroids or individual asteroid families, in more detail. Shape models can also be used in combination with other types of observational data (IR, adaptive optics images, stellar occultations), e.g., to determine sizes and thermal properties.}
   {We use all available photometric data of asteroids to derive their physical models by the lightcurve inversion method and compare the observed pole latitude distributions of all asteroids with known convex shape models with the simulated pole latitude distributions.}
   {We used classical dense photometric lightcurves from several sources (Uppsala Asteroid Photometric Catalogue, Palomar Transient Factory survey, and from individual observers) and sparse-in-time photometry from the U.S.~Naval Observatory in Flagstaff, Catalina Sky Survey, and La Palma surveys (IAU codes 689, 703, 950) in the lightcurve inversion method to determine asteroid convex models and their rotational states. We also extended a simple dynamical model for the spin evolution of asteroids used in our previous paper.}
   {We present 119 new asteroid models derived from combined dense and sparse-in-time photometry. We discuss the reliability of asteroid shape models derived only from Catalina Sky Survey data (IAU code 703) and present 20 such models. By using different values for a scaling parameter $c_{\mathrm{YORP}}$ (corresponds to the magnitude of the YORP momentum) in the dynamical model for the spin evolution and by comparing synthetics and observed pole-latitude distributions, we were able to constrain the typical values of the $c_{\mathrm{YORP}}$ parameter as between 0.05 and 0.6.}
   {}
 
   \keywords{minor planets, asteroids: general - photometry - models}

  \titlerunning{Asteroids' physical models}
  \maketitle

\section{Introduction}\label{introduction}

The lightcurve inversion method (LI) was developed by \citet{Kaasalainen2001a} and \citet{Kaasalainen2001b}. This powerful tool allows us to derive
physical models of asteroids (their rotational states and the shapes) from series of disk-integrated photometry.

\textit{Convex} asteroid shape models can be derived from two different types of disk-integrated photometry: dense or sparse-in-time. Originally, only dense photometry was used. About 20 such dense lightcurves from at least four or five apparitions are necessary for a unique shape determination. By this approach, $\sim100$ asteroid models have been derived \citep[e.g.,][]{Kaasalainen2002b
, Michalowski2004
, Durech2007
, Marciniak2007, Marciniak2008
}. To significantly enlarge the number of asteroid models, sparse photometric data were studied and used in the LI. \citet{Durech2009} determined 24 asteroid models from a combination of dense data with sparse photometry from the U.S. Naval Observatory in Flagstaff (USNO-Flagstaff station, IAU code 689). Sparse data from seven astrometric surveys (including USNO-Flagstaff station) were used in the LI by \citet{Hanus2011}, who presented 80 asteroid models. 16 models were based only on sparse data, the rest on combined dense and sparse data.

Models of asteroids derived by the lightcurve inversion method are stored in the Database of Asteroid Models from Inversion Techniques \citep[DAMIT\footnote{\texttt{http://astro.troja.mff.cuni.cz/projects/asteroids3D}},][]{Durech2010}. In October 2012, models of 213 asteroids were included there.

A larger number of asteroids with derived models of their convex shapes and rotational states is important for further studies. Large statistical samples of physical parameters can tell us more about processes that take place in the asteroids' populations (near-Earth asteroids, main-belt asteroids, or asteroids in individual families). For example, 
an anisotropy of spin-axis directions is present in the population of main-belt asteroids with diameters $\lesssim30$ km \citep{Hanus2011}, where the YORP effect\footnote{Yarkovsky--O'Keefe--Radzievskii--Paddack effect, a torque caused by the recoil force from anisotropic thermal emission, which can alter the rotational periods and orientation of spin axes, see e.g., \citet{Rubincam2000,Vokrouhlicky2003}}, together with 
collisions and mass shedding, is believed to be responsible. There are similar effects on the rotational states of main-belt binaries \citep{Pravec2012}. Convex shape models were also used in combination with 
stellar occultations by asteroids where global nonconvexities can be detected, and the diameter can be estimated with a typical uncertainty of 10\% \citep[see ][]{Durech2011}.

In Section \ref{sec:models}, we describe the dense and sparse photometric data used in the lightcurve inversion method and present new asteroid models derived from combined
photometric data sets or from the sparse-in-time data from the Catalina Sky Survey Observatory (IAU code 703) alone. The reliability tests for derived models are also described.
In Section \ref{sec:YORP_sim}, we use a theoretical model of the latitude distribution of pole directions published in \citet{Hanus2011} in a numerical simulation to constrain the free scaling parameter $c_{\mathrm{YORP}}$ describing our uncertainty in the shape and the magnitude of the YORP momentum.

\section{Asteroid models}\label{sec:models}

We used four main sources of dense photometric lightcurves: 
(i)~the Uppsala Asteroid Photometric Catalogue \citep[UAPC\footnote{\texttt{http://asteroid.astro.helsinki.fi/}}, ][]{Lagerkvist1987, Piironen2001}, where lightcurves for about 1\,000 asteroids are stored, 
(ii)~data from a group of individual observers provided via the Minor Planet Center in the Asteroid Lightcurve Data Exchange Format \citep[ALCDEF\footnote{\texttt{http://www.minorplanet.info/alcdef.html}},][]{Warner2009}, 
(iii)~data from another group of individual observers available online via Courbes de rotation d'ast\' ero\" ides et de com\` etes (CdR\footnote{\texttt{http://obswww.unige.ch/$\sim$behrend/page2cou.html}}), and 
(iv)~data from the Palomar Transient Factory survey \citep[PTF\footnote{\texttt{http://www.astro.caltech.edu/ptf/}},][]{Rau2009}. 
\citet{Polishook2012} recently analyzed a small fraction of PTF data and presented dense lightcurves for 624 asteroids.
So far, only a fraction of photometric data from the PTF has been processed (four overlapping fields on four consecutive nights), which means that this source will become very important in the near future. 

We downloaded sparse data from the AstDyS site (Asteroids -- Dynamic Site\footnote{\texttt{http://hamilton.dm.unipi.it/}}) and gathered sparse lightcurves from the USNO-Flagstaff station (IAU code 689) for $\sim1\,000$ asteroids, from Roque de los Muchachos Observatory, La Palma (IAU code 950) for $\sim500$ asteroids and $\gtrsim100$ sparse data points from the Catalina Sky Survey Observatory \citep[CSS for short, IAU code 703,][]{Larson2003} for $\sim4\,000$ asteroids. We present 119 asteroid models derived from combined dense and sparse data (Section \ref{sec:combined}) and 20 models based only on CSS data (Section \ref{sec:CSS}).

During the model computation, a priori information about the rotational period of the asteroid was used, which significantly reduced the volume of the multidimensional parameter space that had to be searched, and saved computational time. Period values were taken from the regularly updated Minor Planet Lightcurve Database\footnote{\texttt{http://cfa-www.harvard.edu/iau/lists/Lightcurve\-Dat.html}} \citep{Warner2009}. If the period was unknown or insecure, we searched the model over all possible period values of 2--100 hours (usually, when only sparse data are available).

\subsection{Reliability tests}\label{sec:tests}

We carefully tested the reliability of derived models. If we had several dense lightcurves and sparse data from USNO-Flagstaff station for an asteroid, we considered a model as unique if:
(i) the modeled sidereal rotational period was close to the synodic rotational period determined from a single apparition dense data set (synodic period values have usually been previously published and were available in the Minor Planet Lightcurve Database),
(ii) the shape model rotated close to its axis with a maximum momentum of inertia (it was in a relaxed rotational state), and
(iii) models with half and double period values that gave significantly worse fits. 

It was necessary to apply additional tests to models derived from sparse-in-time data alone. We used the tests presented in \citet{Hanus2011} (for more details, see Section 3.3 there), and they were sufficient if photometry from USNO-Flagstaff station was present. In \citet{Hanus2012}, we have shown that reliable asteroid models can also be derived from the Catalina Sky Survey data alone, and we described a convenient procedure for how to proceed during the computation when the rotational period is unknown: the solution should be searched for all periods in an interval of 2--100 hours, and the stability of the solution should be tested for at least two different shape parametrizations\footnote{Shape is represented by coefficients of its expansion into spherical harmonic functions to the order $n$. We call $n$ the shape resolution, the number of shape parameters is then $(n+1)^2$, and our typical value for the shape resolution is $n=6$}. The correct solution had to be stable for both low ($n=3$) and high ($n=6$) shape resolutions. We followed these 
recommendations: we searched for the model in the multidimensional parameter space for shape resolutions $n=3$ and $n=6$ and checked that we derived solutions with similar rotational states. In \citet{Hanus2012}, we tested values $n=2,3,4,5,6$ for the shape resolution. Correct solutions (i.e., models from the CSS data were similar to the models based on different data sets) were reproduced for most values of $n$. On the other hand, incorrect solutions were derived only for values $n=6$ and sometimes also for $n=4$ or $n=5$, but never for $n=2$ or $n=3$.

\subsection{Models from combined dense and sparse data}\label{sec:combined}

The shape model determination scheme was very similar to the one used in \citet{Hanus2011}. 119 new asteroid models were derived because we gathered $\sim1000$ new dense lightcurves from ALCDEF, another $\sim1000$ lightcurves from PTF, $\sim$300 from individual observers, and also additional sparse data observed by the CSS during the second half of the year 2010 and the first half of the year 2011. Derived rotational states with basic information about the photometry used for 119 asteroids are listed in Table \ref{tab:combined}. Out of them, 18 models are based only on {\em combined} sparse data from various sources, but in all cases, sparse data from USNO-Flagstaff station were present\footnote{Models based only on data from the Catalina Sky Survey and described later in Section~\ref{sec:CSS}}. In Table~\ref{tab:references}, we list the references to the dense lightcurves we used for the new model determination.

Although the amount of photometric data from PTF was similar to that from ALCDEF, only two new shape models (for asteroids with numbers 52\,820 and 57\,394, see Table~\ref{tab:combined}) were derived with their contribution. The first reason was a significantly worse quality of PTF data: only for 84 asteroids out of 624 were the data sufficient for determining a synodic period, while other lightcurves were noisy or burdened with systematic errors. In many cases they allowed only for an estimate of a lower limit for the lightcurve amplitude \citep[presented in][]{Polishook2012}. The second reason was that PTF data alone were not sufficient for a unique model determination (they covered only one apparition), no other dense lightcurves were usually available, and sparse data were available for only fewer than a half of these asteroids. Many asteroids detected by the PTF survey were previously unknown. 

There are previously published models available for 15 of the asteroids modeled 
here: (11) Parthenope, (79) Eurynome, (272) Antonia, (281) Lucretia, 
(351) Yrsa, (352) Gisela, (390) Alma, (787) Moskva, (852) Wladilena, (1089) Tama, 
(1188) Gothlandia, (1389) Onnie, (1572) Posnania, (1719) Jens, and (4954) Eric 
(see databases by \citealt{Kryszczynska2007} and \citealt{Warner2009}). As these models were usually based on limited datasets, 
our solutions differ from some of them substantially, while agreeing  
for some in the spin axis latitude or the sidereal period value. We fully
confirmed previous models for six objects of that sample: the spin models 
of (79) Eurynome by \citet{Michalowski1996b}, (787) Moskva by 
\citet{Svoren2009}, and (1572) Posnania by \citet{Michalowski2001},
as well as our preliminary solutions for (390) Alma, (1389) Onnie, and (1719) Jens obtained in 
\citet{Hanus2011}. 

The shape models and their spin solutions can be 
found in the DAMIT database \citep{Durech2010}. We noticed that 
for the models based only on sparse data, their shapes tend to be very 
angular, with sharp edges and large planar areas, thus can be treated 
only as crude approximations of the real asteroid shapes. However, a substantial 
addition ($\gtrsim10$ lightcurves from $\gtrsim2$ apparitions) of dense lightcurves smooths the shape models out, making 
them look more realistic, as confirmed by their better fit to 
occultation chords.

\begin{figure}
	\begin{center}
	 \resizebox{\hsize}{!}{\includegraphics{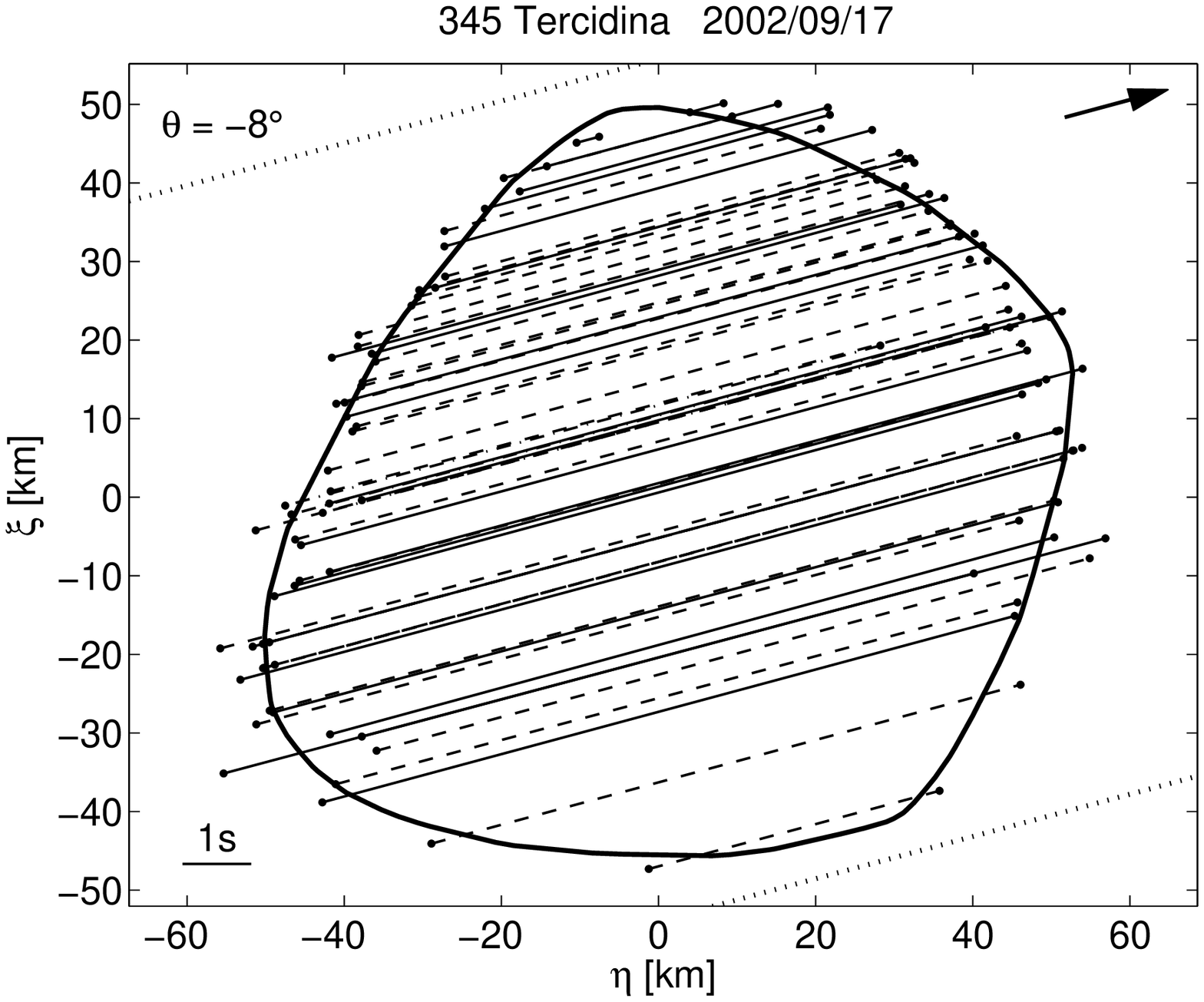}\includegraphics{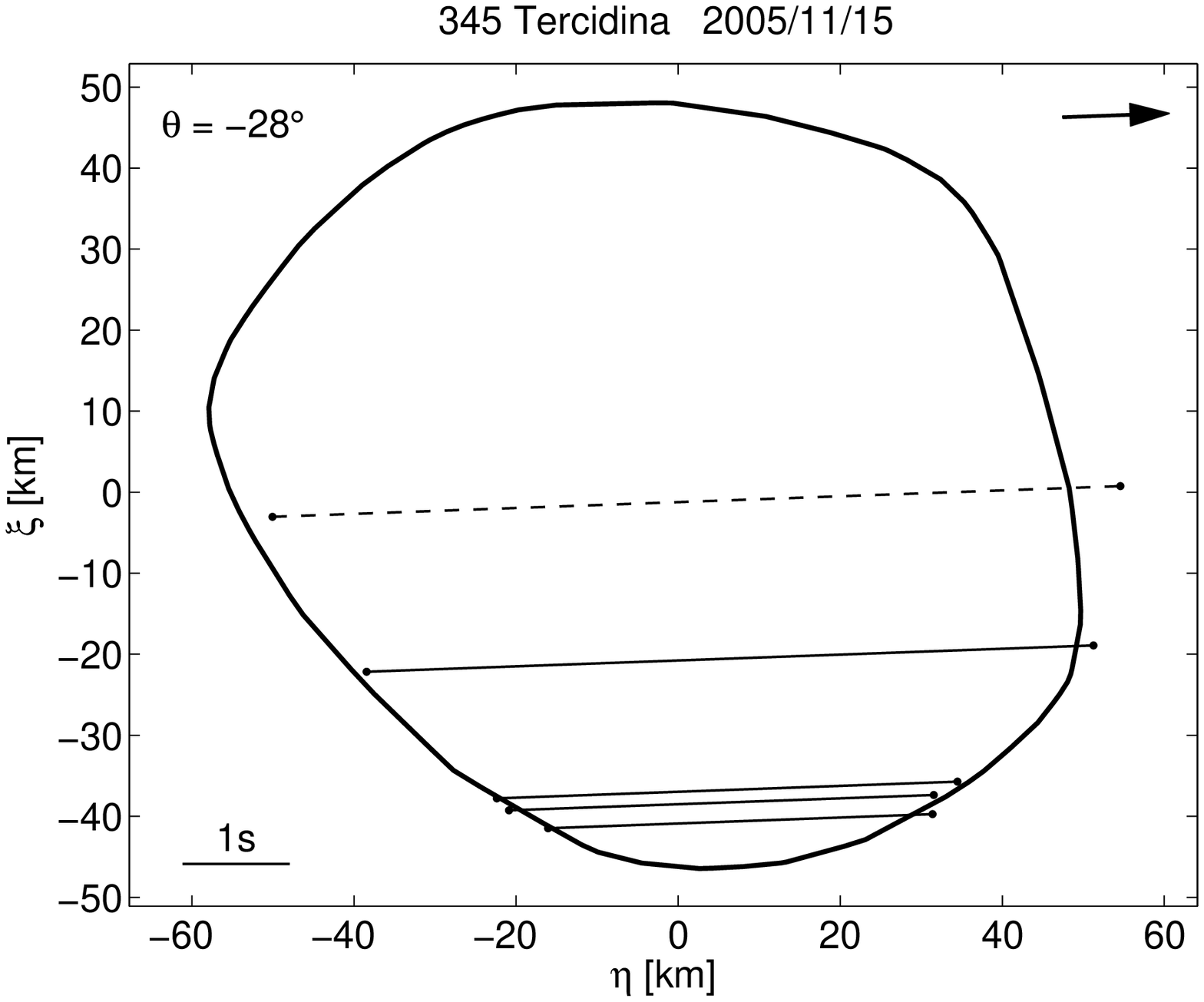}}\\
	\end{center}
	\caption{\label{fig:occ_345}Two observations of star occultations by asteroid (345) Tercidina. The solid contour corresponds to a scaled projected silhouette of the shape model with the pole $(346^{\circ}, $--$55^{\circ})$, each chord represents one occultation observation (solid lines are CCD, video, or photoelectric observations; dashed lines are visual observations, and dotted lines negative observations). Each plot also contains the time scale (lower left corner), the latitude of the sub-Earth point $\theta$ for the time of occultation (upper left corner), and the direction of the relative velocity (the arrow in the upper right corner). East points to the left and north up.}
\end{figure}

\begin{figure}
	\begin{center}
	 \resizebox{\hsize}{!}{\includegraphics{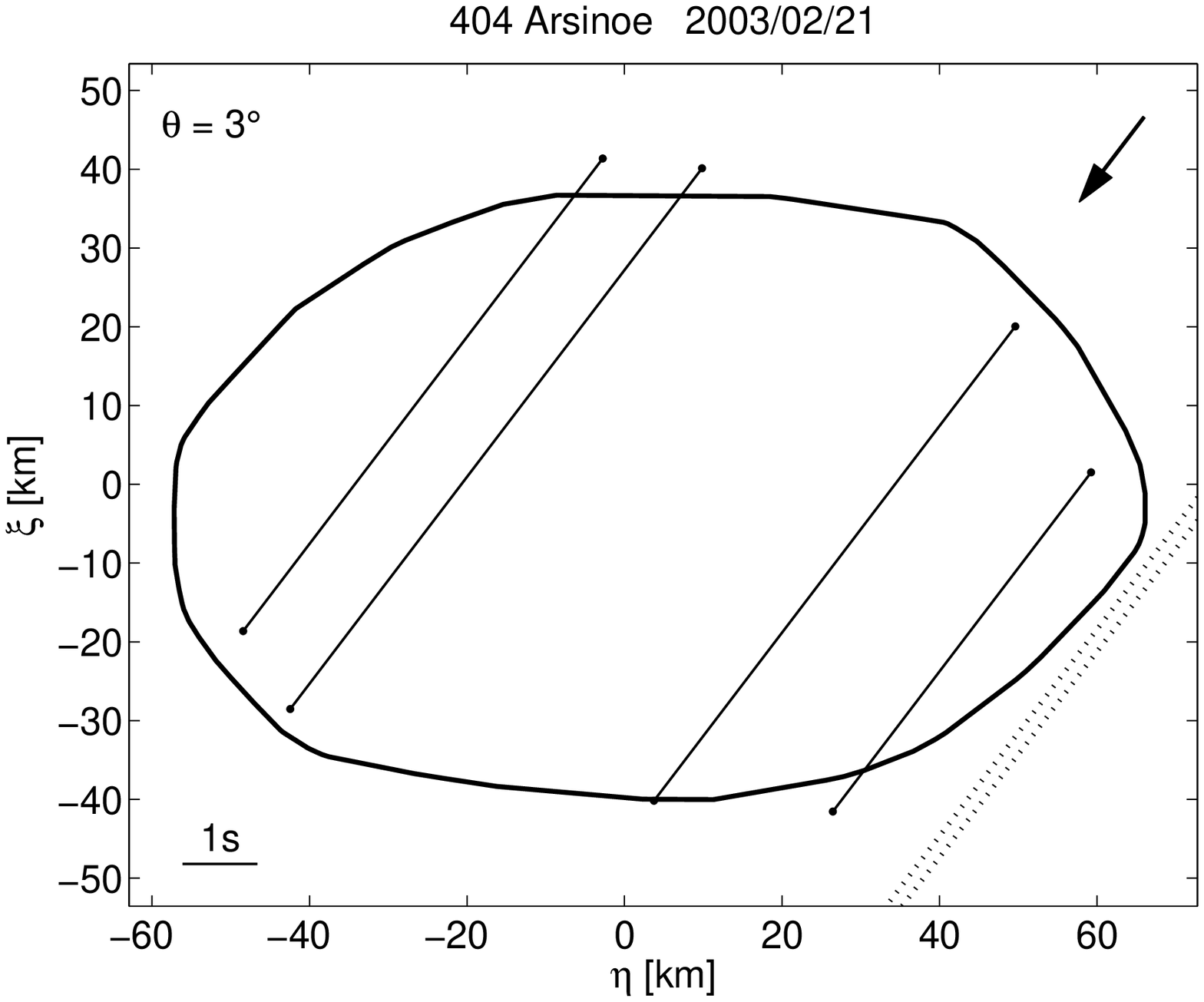}\includegraphics{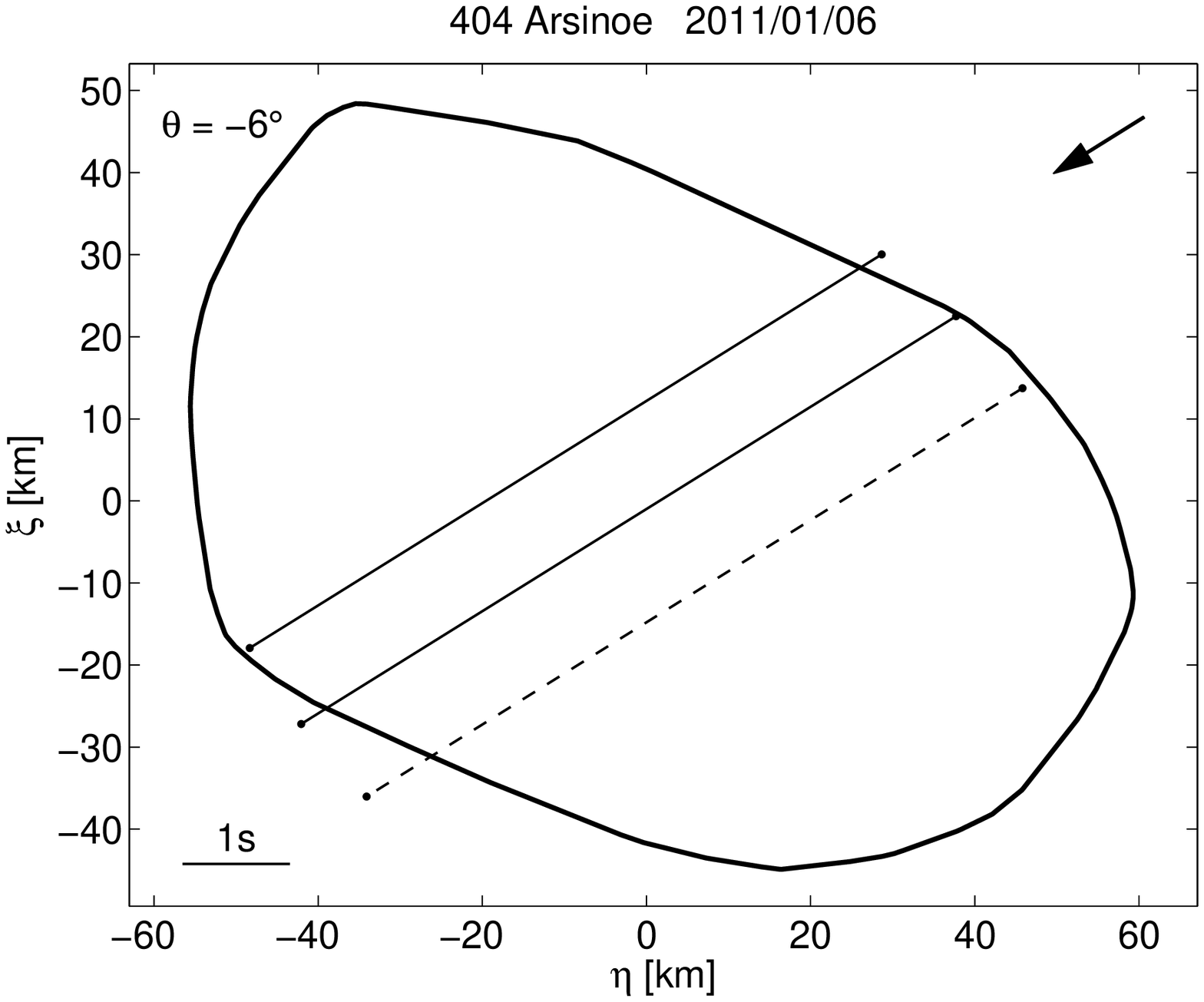}}\\
	\end{center}
	\caption{\label{fig:occ_404}Two observations of star occultations by asteroid (404) Arsinoe. The solid contour corresponds to a scaled projected silhouette of the shape model with the pole $(25^{\circ}, 57^{\circ})$. See Fig.~\ref{fig:occ_345} for line types and symbols explanation.}
\end{figure}

\begin{figure}
	\begin{center}
	 \resizebox{\hsize}{!}{\includegraphics{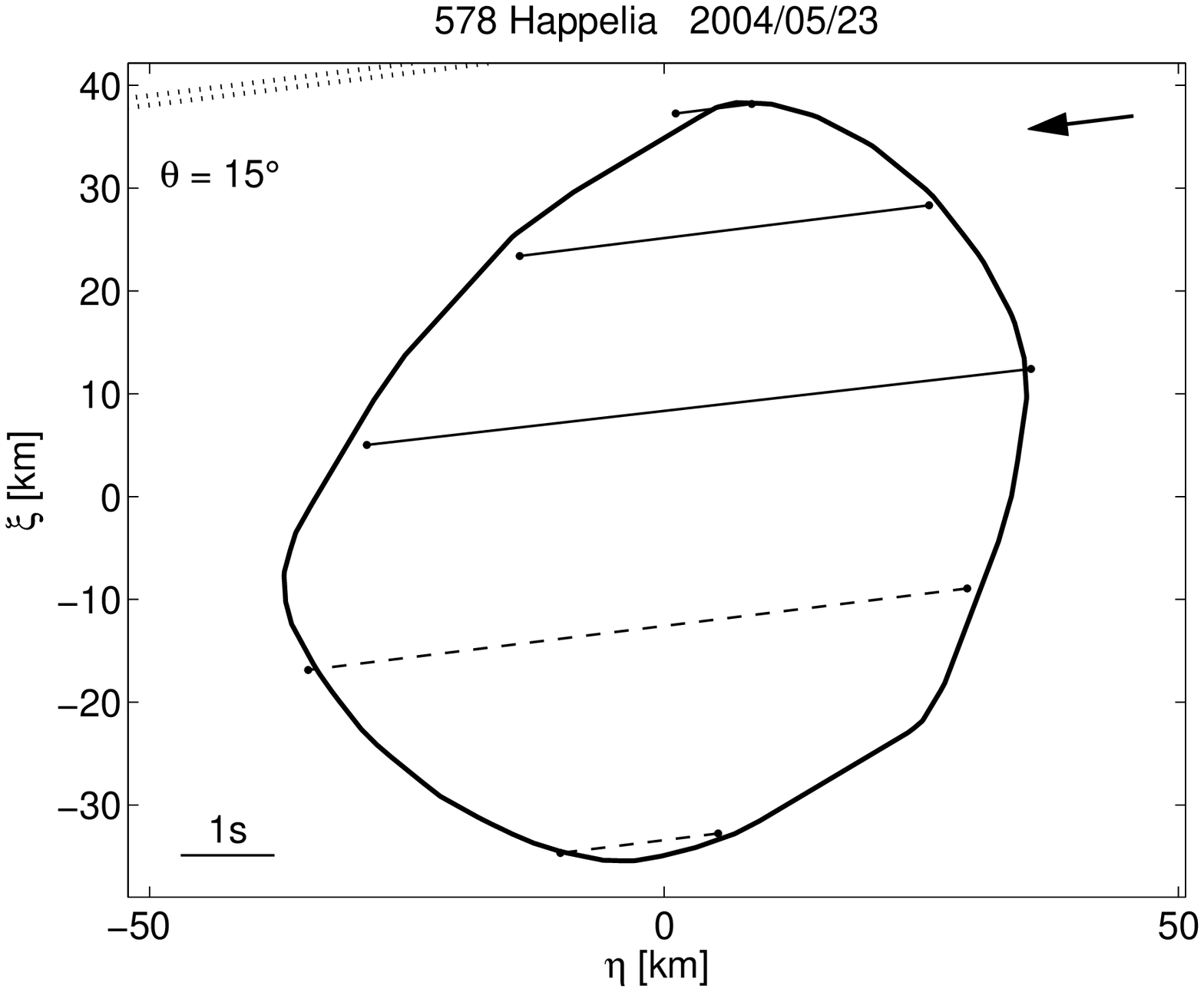}\includegraphics{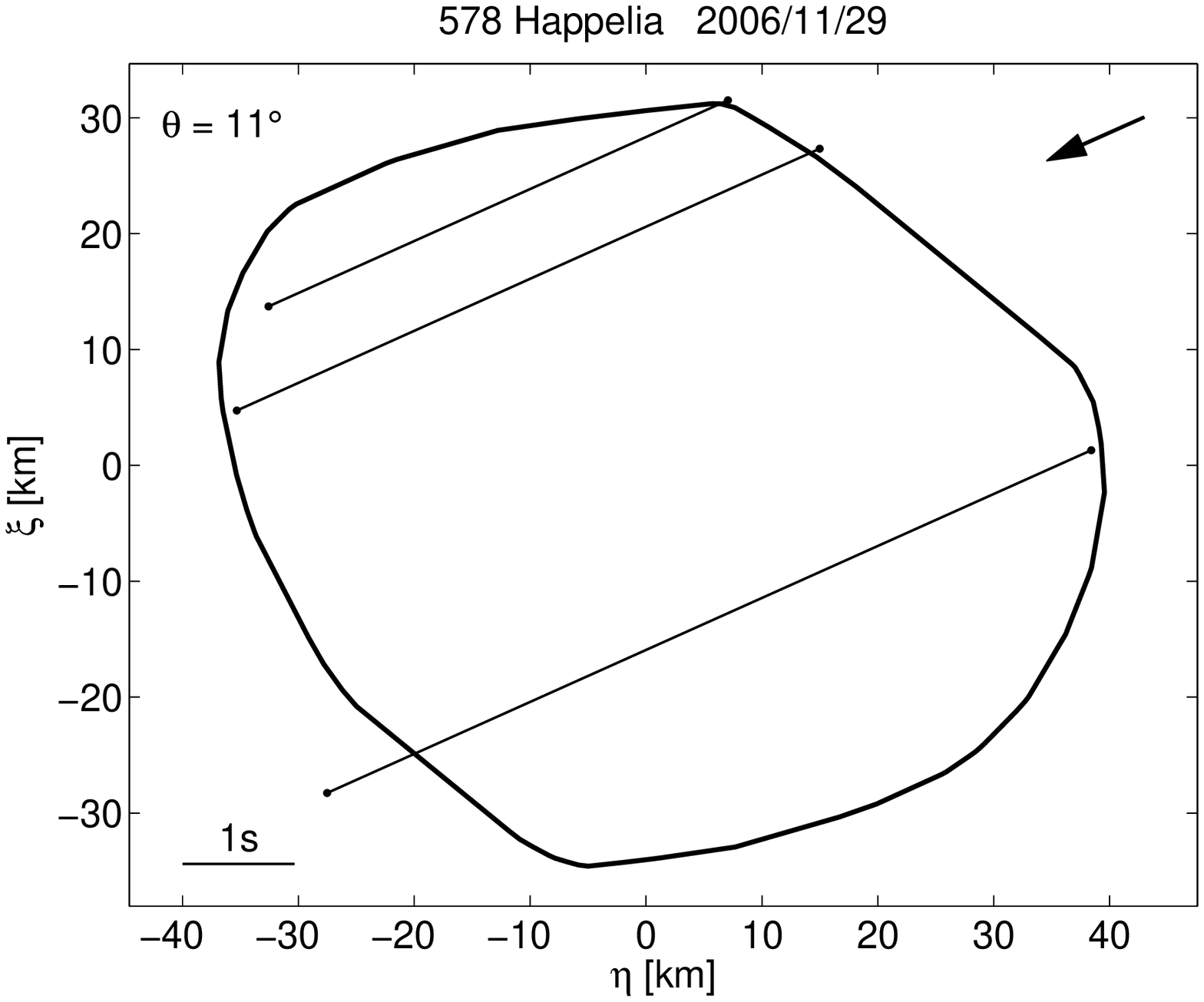}}\\
	\end{center}
	\caption{\label{fig:occ_578}Two observations of star occultations by asteroid (578) Happelia. The solid contour corresponds to a scaled projected silhouette of the shape model with the pole $(339^{\circ}, 62^{\circ})$. See Fig.~\ref{fig:occ_345} for line types and symbols explanation.}
\end{figure}

From observations of star occultations by asteroids, we can reconstruct asteroid projected silhouettes. These silhouettes can then be compared with the predicted contours of the convex shape models and used for the asteroid size determination by scaling the shape models to fit the occultation chords. A reasonable number of observations were available for three asteroids from our sample. By using the same methods as in \citet{Durech2011}, we rejected mirror solutions for the asteroids (345)~Tercidina and (578)~Happelia, and also determined equivalent diameters (corresponding to spheres with the same volume as the scaled convex shape models): 96$\pm$10~km for (345)~Tercidina, 101$\pm$5~km for (404)~Arsinoe, and 70$\pm$5~km for (578)~Happelia. Two different stellar occultations are available for all three asteroids, and are plotted in Figs.~\ref{fig:occ_345},~\ref{fig:occ_404},~and~\ref{fig:occ_578}.

During the apparition in 2004, the lightcurves of asteroid (1089)~Tama have shown features typical of close binary systems \citep{Behrend2004} and indeed, the system was later interpreted as a synchronous close binary \citep{Behrend2006}. Our {\em brick-like} convex shape model is strongly elongated with sharp edges and is similar to a convex shape model of a close binary system (90)~Antiope. Such a shape appearance for close binaries was predicted from synthetic data \citep{Durech2003}. 

\subsection{Models based on data from the Catalina Sky Survey astrometric project}\label{sec:CSS}

There are two different groups of asteroid models based on CSS data: 
(i) models with previously reported synodic periods determined from dense data (we did not have these dense data, so period values were taken from the literature, usually from the Minor Planet Lightcurve Database), and
(ii) models with previously unknown rotational periods.
In the first case, we could compare the published period value with the period value derived by the LI (see Table \ref{tab:CSS}, columns 7 and 9). If both periods agreed within their uncertainties, we considered the solution reliable. This test could not be performed for the second group of models, so we had to use additional reliability tests (see Section \ref{sec:tests}). 

In Table~\ref{tab:CSS}, we present 20 asteroid models based only on the CSS data. The previous period estimates were not available for 12 of them. All of these 20 models have higher uncertainties of the pole orientations and lower shape resolution than models based on combined data, and all are possible candidates for follow-up lightcurve observations for period confirmation and more detailed shape determination.

\section{Semi-empirical scaling of the YORP effect}\label{sec:YORP_sim}

\begin{figure}
	\begin{center}
	 \resizebox{\hsize}{!}{\includegraphics{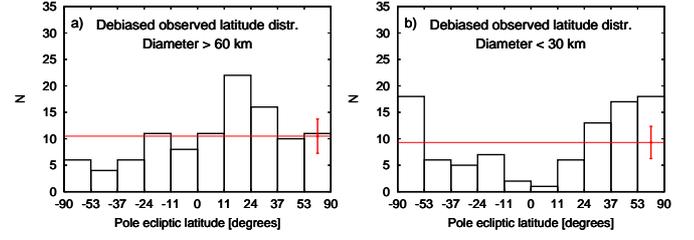}}\\
	\end{center}
	\caption{\label{fig:latitude_distribution}Debiased observed latitude distribution of main-belt asteroids with diameters $D>60$~km (left panel) and $D<30$~km (right panel). The latitude bins are equidistant in sin~$\beta$. The thin horizontal line corresponds to the average value $\bar N$ and the errorbar to $\sqrt{\bar N}$.}
\end{figure}

Our enlarged sample of physical parameters for $\sim$330 asteroids\footnote{According to the asteroid size distribution function of \citet{Davis2002}, we have in our sample $\sim$ 30\% of all asteroids with $D>100$~km, $\sim$ 15\% asteroids with $60$~km~$<D<100$~km, and $\sim$ 14\% asteroids with $30$~km~$<D<60$~km} validates our previous results based on a smaller asteroid sample (220 asteroids) presented in \citet{Hanus2011}. In Fig.~\ref{fig:latitude_distribution}, we show the observed {\em debiased} \citep[i.e., we removed the systematic effect of the lightcurve inversion method caused by the method having a higher probability of deriving a unique solution for asteroids with larger pole latitudes. The debiasing procedure was based on a numerical simulation presented in][see Section 4.3 there]{Hanus2011} latitude distributions of pole directions for main-belt asteroids with diameters $D<30$~km and $D>60$~km. The population of larger asteroids ($D>60$~km) exhibits an excess of prograde rotators, probably of primordial origin \citep[predicted also from numerical simulations by][]{Johansen2010}. On the other hand, smaller asteroids ($D<30$~km) have a clearly bimodal latitude  distribution -- most of the asteroids have ecliptic pole latitudes $>53^{\circ}$. 

The debiased observed latitude distribution of the pole directions of MBAs represents fingerprints from the past evolution of this population. Direct comparison between the observed asteroid properties and predictions of theoretical models can validate/exclude some of the asteroid dynamical evolution theories or constrain specific free parameters.

In \citet{Hanus2011}, we introduced a simple dynamical model for the spin evolution of asteroids, where we included (i)~the YORP thermal effect, (ii)~random reorientations induced by noncatastrophic collisions, (iii)~oscillations caused by gravitational torques and spin-orbital resonances, and also (iv)~mass shedding when a critical rotational frequency is reached. Because we studied a large statistical sample of asteroids, the effect on the overall latitude distribution of pole directions caused by other processes (gravitational torques by the Sun, damping, or tumbling) was assumed to be only minor.

The model was based on the relations for the rate of the angular velocity $\omega$ ($\omega=2\pi/P$) and the obliquity $\epsilon$ (Euler equations)
\begin{eqnarray}
{{\rm d}\omega\over{\rm d} t} &=& c f_i(\epsilon)\,,\qquad i = 1 \dots
200\,,\label{eq:domega}\\
{{\rm d}\epsilon\over{\rm d} t} &=&
{c g_i(\epsilon)\over\omega}\,,\label{eq:depsil}
\end{eqnarray}
where $f$- and $g$-functions describing the YORP effect for a set of 200 shapes with the effective radius $R_0 = 1\,{\rm km}$, the bulk density $\rho_0 = 2500\,{\rm kg}/{\rm m}^3$, located on a circular orbit with the semi-major axis $a_0 = 2.5\,{\rm AU}$, were calculated numerically by \citet{Capek2004}. We assigned one of the artificial shapes (denoted by the index~$i$) for each individual asteroid from our sample\footnote{We did not use the convex-hull shape models derived in this work because the two samples of shapes are believed to be statistically equivalent, and moreover, the YORP effect seems sensitive to small-scale surface structure \citep{Scheeres2007}, which cannot be caught by our shape models.}.
The $f$- and $g$-functions were scaled by a factor
\begin{equation}\label{cyorp}
c = c_{\rm YORP} \left({a\over a_0}\right)^{-2} \left(R\over R_0\right)^{-2}
\left(\rho_{\rm bulk}\over\rho_0\right)^{-1}\,,
\end{equation}
where $a$, $R$, $\rho_{\rm bulk}$ denote the semi-major axis, the radius, and the density of the simulated body, respectively, and $c_{\rm YORP}$ is a {\em free} scaling parameter reflecting our uncertainty in the shape models and the magnitude of the YORP torque, which dependents on small-sized surface features \citep[even boulders,][]{Statler2009} and other simplifications in the modeling of the YORP torque.

We enhanced the simulation of the spin evolution of asteroids presented in \citet{Hanus2011}, by testing different values of the free parameter $c_{\rm{YORP}}$ and comparing the resulting synthetic latitude distributions with the observed debiased latitude distributions. Thanks to the new asteroid models, we had an updated observed spin vector distribution. We added 50\% more observed asteroids, so we used 307 instead of 220 models for this comparison. 

We used the following values of the parameter $c_{\rm{YORP}}$: 0.01, 0.05, 0.1, 0.2, 0.3, 0.4, 0.5, 0.6, 0.8. Values of $c_{\rm{YORP}} \gtrsim 1$ were already recognized as unrealistic.

For each value of $c_{\rm YORP}$, we ran 100 simulations with different random seeds to generate different initial $\omega$ and spin vector distributions. We integrated Eqs.~(\ref{eq:domega}) and (\ref{eq:depsil}) numerically. The time span was 4 Gyr with the time step of the explicit Euler scheme $\Delta t=10$~Myr. As initial conditions, we assumed a Maxwellian distribution of angular velocities $\omega$ and isotropically distributed spin vectors. We also used $K = 10^{-2}\,{\rm W}/{\rm K}/{\rm m}$, $\rho_{\rm bulk} = 2500\,{\rm kg}/{\rm m}^3$.

Every time a critical angular velocity ($\omega_{\rm crit} = \sqrt{4/3 \pi G \rho_{\rm bulk}}$) was reached for an asteroid, we assumed a mass shedding event, so that we reset the rotational period to a random value from an interval of 2.5,9 hours. We altered the assigned shape, but we kept the sense of the rotation and the orientation of the spin axis. We also included a simple Monte-Carlo model for the spin axis reorientations caused by collisions \citep[with $\tau_{\rm reor} = B \left({\omega\over\omega_0}\right)^{\beta_1} \left({D \over D_0}\right)^{\beta_2}$, where $B = 84.5\,{\rm kyr}$, $\beta_1 = 5/6$, $\beta_2 = 4/3$, $D_0 = 2\,{\rm m}$, and $\omega_0$ corresponds to period $P = 5$~hour,][]{Farinella1998}. After the collision, we reset the spin axis and period to random values (new period was from an interval of 2.5,9~hours). Collisional disruptions are not important in our case so they were not considered. We also accounted for spin-orbital resonances by adding a sinusoidal oscillation to $\beta$ \citep[to prograde rotators, only,][]{Vokrouhlicky2006} with a random phase and an amplitude $\simeq 40^\circ$.

The spin states of our synthetic asteroids evolve during the simulation. At each time $t$ of the simulation, we can construct a latitude distribution of the pole directions with the latitude values split into ten bins with a variable width corresponding to constant surface on the celestial sphere. Because we used ecliptic coordinates with the longitude $\lambda$ and the latitude $\beta$, the bins were equidistant in sin\,$\beta$. To describe the temporal evolution of the simulated latitude distributions, we computed a $\chi^2$ metric between the simulated and the debiased observed latitude distributions of asteroids with diameters $D<60$~km. The assumption of isotropically distributed initial spin vectors is not fulfilled for larger asteroids ($D>60$~km), because this population has an excess of prograde rotators (see Fig.~\ref{fig:latitude_distribution}), which is believed to have a primordial origin \citep{Johansen2010}. The second reason we rejected asteroids with $D>60$~km from latitude comparison is that their evolution is rather slow compared to the simulation time span.

For each time $t$ within the simulation run $j$ ($j=1..100$), the corresponding chi-square value $\chi^2_{tj}$ is defined by
\begin{equation}
\chi^2_{tj} \equiv \sum_i\frac{(S_{tji} - O_i)^2}{\sigma^{2}_{tji}},
\end{equation}
where $S_{tji}$ denotes the number of synthetic bodies with latitudes in bin $i$, $O_i$ the number of observed latitudes in bin $i$, and $\sigma_{tji}\equiv\sqrt{S_{tji}+O_i}$ corresponds to the uncertainty estimate.

\begin{figure}
	\begin{center}
	 \resizebox{\hsize}{!}{\includegraphics{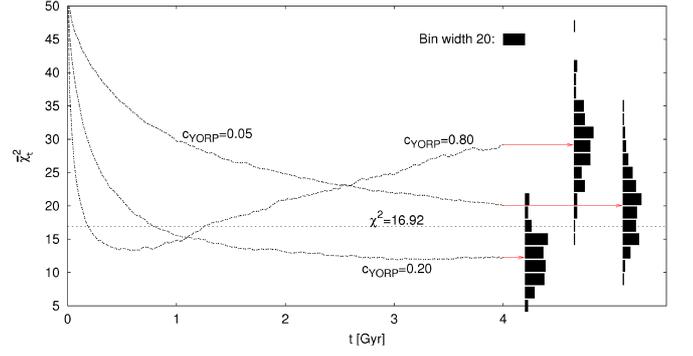}}\\
	\end{center}
	\caption{\label{fig:chi_vs_time}Temporal evolution of the ${\chi}^2$ that corresponds to the difference between the simulated latitude distributions, averaged over all 100 runs, and the debiased observed latitude distribution (i.e., $\bar{\chi}^2_t$) for three different values of parameter $c_{\rm YORP}$ = 0.05, 0.20, and 0.80 (we performed a chi-square test). Vertical histograms on the righthand side represent the distributions of $\chi^2_{tj}$ at time $t=4$ Gy for all 100 runs. Dotted line: the statistically significant probability value of 5\%, i.e. $\chi^2=16.92$.}
\end{figure}

In Fig.~\ref{fig:chi_vs_time}, we show the temporal evolution of the {\em average} chi-square $\bar{\chi}^2_t=\sum_j\chi^2_{tj}/100$ in the course of our numerical simulations for different $c_{\rm YORP}$ values. As we see in Fig.~\ref{fig:chi_vs_time}, the average synthetic latitude distribution evolves in course of the time (while the debiased observed latitude distribution is fixed). We can distinguish three basic cases of the temporal evolution:
\begin{itemize}
 \item When the YORP effect is weak ($c_{\rm YORP}\lesssim0.1$), the synthetic latitude distribution only evolves slowly and is never similar to the observed latitude distribution, even at the end of the simulation, because $\bar{\chi}^2_t$ is still large (for $N=9$, a statistically significant probability value of 5\% corresponds to $\chi^2=16.92$).
 \item A steady state (i.e., the state when the synthetic latitude distribution does not significantly evolve in time, and thus the $\bar{\chi}^2_t$ is approximately constant) is only reached for $c_{\rm YORP}$ values close to 0.2.
 \item For values $c_{\rm YORP}\gtrsim 0.3$, the synthetic latitude distribution evolves faster and, at a certain time, is most similar to the observed latitude distribution (i.e., the minimum of $\bar{\chi}^2_t$). After that, the $\bar{\chi}^2_t$ grows, because the YORP also significantly develops also larger asteroids, and thus the bins with low latitudes are depopulated more than is observed.
\end{itemize}

\begin{figure}
	\begin{center}
	 \resizebox{\hsize}{!}{\includegraphics{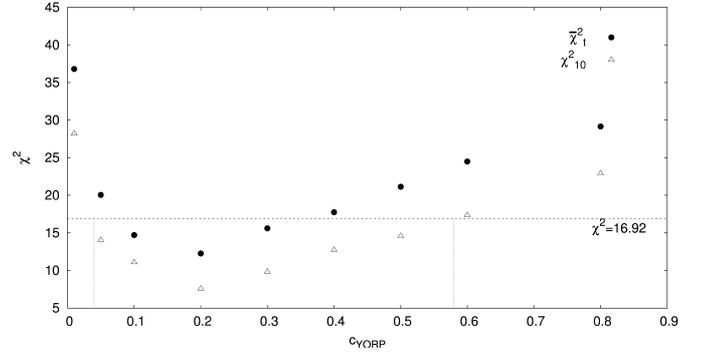}}\\
	\end{center}
	\caption{\label{fig:chi_vs_cyorp}Dependence of $\bar{\chi}^2_t$ and $\chi^2_{10}$ values calculated for the time $t=4$~Gyr (i.e. the final state of the simulation) on different values of the $c_{\rm YORP}$ parameter. We also plotted the statistically significant probability value of 5\% which corresponds to $\chi^2=16.92$ and the interval of plausible $c_{\rm YORP}$ values from 0.05 to 0.6.}
\end{figure}

Vertical histograms on the righthand side of Fig.~\ref{fig:chi_vs_time} represent the distributions of $\chi^2_{tj}$ at the time $t=4$ Gy for all 100 runs. The average chi-square $\bar{\chi}^2_t$ of the model with $c_{\rm YORP}=0.05$ is substantially higher than 16.92, so this model can be considered wrong. However, from the distributions of $\chi^2_{tj}$ we can see that about 25~\% of individual runs have $\chi^2_{tj}$ lower than 16.92. To avoid rejecting those $c_{\rm YORP}$ values that are partially compatible with the observations, we should instead use a more representative value of $\chi^2$ than the average $\bar{\chi}^2_t$, namely a value $\chi^2_{10}$, for which 10\% runs have lower $\chi^2_{tj}$ (see Fig.~\ref{fig:chi_vs_cyorp}). Based on the $\chi^2_{10}$, the most probable values of the $c_{\rm YORP}$ parameter are between 0.05 and 0.6.

\section{Discusion \& conclusions}

Our preferred interpretation of the optimal $c_{\mathrm{YORP}}$ value being much lower than one is that small-scale features (boulders) tend to decrease the YORP torque. This hypothesis is supported by the independent modeling of \citet{Rozitis2012}, who estimate, by including rough surface thermal-infrared beaming effects in their long-term spin evolution model, that the surface roughness is on average responsible for damping the magnitude of the YORP effect typically by half of the smooth surface predictions. This would correspond to $c_{\mathrm{YORP}}=0.5$ in our notation. The YORP effect is sensitive to the sizes of the boulders and can vary tens of percent, so the results of \citet{Rozitis2012} agree with our model.

As an important application, we mention that the constraint for the value of $c_{\mathrm{YORP}}$ can be used in simulations of the long-term dynamical evolution of asteroid families. So far, $c_{\mathrm{YORP}}$ has been used as a free parameter \citep[e.g., in the method presented by][]{Vokrouhlicky2006b}. Constraining $c_{\mathrm{YORP}}$ therefore {\em removes} one free parameter from the simulations and should thus lead to a better determination of the ages of asteroid families.

\vspace{3mm}

Finally, the results of this paper can be briefly summarized as follows.
\begin{itemize}
 \item For 119 asteroids, we derived the convex shape models and rotational states from their combined disk-integrated dense and sparse photometric data. This effort was achieved with the help of $\sim100$ individual observers who were willing to share their lightcurves. The typical uncertainty of the sidereal rotational period is $\sim10^{-5}$ h and of the pole direction 10--20$^{\circ}$. All new models are now included in the DAMIT database.
 \item We also derived 20 asteroid models based purely on sparse-in-time photometry from the Catalina Sky Survey Observatory. The reliability of these models is supported by the fact that for eight of them, we obtained similar rotational period values that were previously reported in the literature and derived from an independent data set (dense photometry). We do not have any previous information about the rotational periods for the 13 other asteroids. Due to relatively larger uncertainties of the CSS sparse data, the typical uncertainty of the sidereal rotational period is $\sim10^{-4}-10^{-5}$ h and of the pole direction 20--40$^{\circ}$.
 \item By combining observations of stellar occultations by asteroids with derived convex shape models, we determined equivalent diameters for the asteroids (345) Tercidina, (404) Arcinoe, and (578) Happelia to 96$\pm$10~km, 101$\pm$5~km and 70$\pm$5~km, respectively.
 \item We updated a simple dynamical model for the spin evolution of asteroids and compared the synthetic pole latitude distributions to the debiased observed latitude distributions of 307 asteroids. By using several values of the scaling parameter $c_{\mathrm{YORP}}$ defined by Eq.~\ref{cyorp} (from 0.01 to 0.8), we constrained its value to $c_{\mathrm{YORP}}\in[0.05, 0.6]$. We interpreted the low value of $c_{\mathrm{YORP}}$ as a result of the surface roughness.
\end{itemize}

\begin{acknowledgements}
The work of JH has been supported by grant GA\,UK 134710 of the Grant
agency of the Charles University and by the project SVV 265301 of the Charles
University in Prague. The work of JH and J\v D has been supported by grants GACR
209/10/0537 and P209/12/0229 of the Czech Science Foundation, the work of JD and MB by the Research
Program MSM0021620860 of the Czech Ministry of Education, and the work of MB also by the grant
GACR 13-01308S of the Grant Agency of the Czech Republic.

The work of TSR was carried out through the Gaia Research for
European Astronomy Training (GREAT-ITN) network. He has 
received funding from the European Union Seventh Framework Program 
(FP7/2007-2013) under grant agreement no.264895.
            
This work is partially based on observations made at
the South African Astronomical Observatory (SAAO).

It was based on observations made with the Nordic Optical Telescope, operated
on the island of La Palma jointly by Denmark, Finland, Iceland,
Norway, and Sweden, in the Spanish Observatorio del Roque de los
Muchachos of the Instituto de Astrofisica de Canarias.  

This work is partially based on observations carried out
with the Pic du Midi Observatory 0.6 m telescope, a facility operated
by the Observatoire Midi-Pyr\'{e}n\'{e}es and Association T60, an amateur
association.

The calculations were performed on the computational cluster Tiger
at the Astronomical Institute of Charles University in Prague (\texttt{http://sirrah.troja.mff.cuni.cz/tiger}). 
\end{acknowledgements}

\bibliography{mybib}
\bibliographystyle{aa}

\onecolumn

\longtab{1}{
\begin{longtable}{r@{\,\,\,}l rrrr D{.}{.}{6} rrrrrc}
\caption{\label{tab:combined}List of new asteroid models derived from combined dense and sparse data or from sparse data alone.}\\
\hline 
\multicolumn{2}{c} {Asteroid} & \multicolumn{1}{c} {$\lambda_1$} & \multicolumn{1}{c} {$\beta_1$} & \multicolumn{1}{c} {$\lambda_2$} & \multicolumn{1}{c} {$\beta_2$} & \multicolumn{1}{c} {$P$} & $N_{\mathrm{lc}}$ & $N_{\mathrm{app}}$  & $N_{\mathrm{689}}$ & $N_{\mathrm{703}}$ & $N_{\mathrm{950}}$ \\
\multicolumn{2}{l} { } & [deg] & [deg] & [deg] & [deg] & \multicolumn{1}{c} {[hours]} &  &  &  &  & \\
\hline\hline

\endfirsthead
\caption{continued.}\\

\hline
\multicolumn{2}{c} {Asteroid} & \multicolumn{1}{c} {$\lambda_1$} & \multicolumn{1}{c} {$\beta_1$} & \multicolumn{1}{c} {$\lambda_2$} & \multicolumn{1}{c} {$\beta_2$} & \multicolumn{1}{c} {$P$} & $N_{\mathrm{lc}}$ & $N_{\mathrm{app}}$  & $N_{\mathrm{689}}$ & $N_{\mathrm{703}}$ & $N_{\mathrm{950}}$ \\
\multicolumn{2}{l} { } & [deg] & [deg] & [deg] & [deg] & \multicolumn{1}{c} {[hours]} &  &  &  &  & \\
\hline\hline
\endhead
\hline
\endfoot
11 & Parthenope & 311 & 14 & 128 & 14 & 13.72205 & 107 & 13 & 297 & 24 & 147 \\
25 & Phocaea & 347 & 10 &  &  & 9.93540 & 22 & 5 & 272 & 100 &  \\
72 & Feronia & 287 & $-$39 & 102 & $-$55 & 8.09068 & 20 & 5 & 196 & 124 & 127 \\
79 & Eurynome & 228 & 30 & 54 & 24 & 5.97772 & 36 & 4 & 240 & 168 &  \\
147 & Protogeneia & 269 & 15 & 90 & 14 & 7.85232 & 11 & 3 & 152 & 80 &  \\
149 & Medusa & 333 & $-$73 & 156 & $-$76 & 26.0454 & 13 & 4 & 134 & 60 &  \\
157 & Dejanira & 319 & $-$64 & 146 & $-$33 & 15.8287 & 14 & 2 & 94 & 123 &  \\
166 & Rhodope & 345 & $-$22 & 173 & $-$3 & 4.714793 & 7 & 2 & 141 & 111 &  \\
178 & Belisana & 260 & 20 & 79 & 9 & 12.32139 & 35 & 3 & 147 & 127 &  \\
183 & Istria & 85 & 20 &  &  & 11.76897 & 8 & 2 & 142 & 174 &  \\
193 & Ambrosia & 141 & $-$11 & 328 & $-$17 & 6.58166 & 18 & 4 & 169 & 87 &  \\
199 & Byblis & 344 & $-$24 & 165 & 9 & 5.22063 & 22 & 5 & 184 & 108 & & \\
220 & Stephania & 26 & $-$50 & 223 & $-$62 & 18.2087 & 9 & 2 & 117 & 99 &  \\
222 & Lucia & 106 & 50 & 293 & 49 & 7.83671 & 9 & 4 & 160 & 100 &  \\
242 & Kriemhild & 100 & $-$40 & 285 & $-$15 & 4.545174 & 25 & 7 & 179 & 144 &  \\
257 & Silesia & 5 & $-$53 & 176 & $-$46 & 15.7097 & 18 & 2 & 167 & 88 &  \\
260 & Huberta & 23 & $-$28 & 206 & $-$19 & 8.29055 & 6 & 2 & 162 & 90 &  \\
265 & Anna & 109 & $-$53 &  &  & 11.6903 &  &  & 114 & 79 &  \\
272 & Antonia & 293 & $-$90 &  &  & 3.85480 & 7 & 2 & 109 & 92 &  \\
281 & Lucretia & 128 & $-$49 & 309 & $-$61 & 4.349711 & 8 & 4 & 129 & 83 & \\
290 & Bruna & 286 & $-$80 & 37 & $-$74 & 13.8055 & 9 & 1 & 97 & 66 &  \\
297 & Caecilia & 223 & $-$53 & 47 & $-$33 & 4.151388 & 15 & 5 & 149 & 130 &  \\
345 & Tercidina & 346 & $-$55 &  &  & 12.37082 & 42 & 8 & 161 & 155 & \\
351 & Yrsa & 20 & $-$70 & 193 & $-$41 & 13.3120 & 2 & 1 & 183 & 52 &  \\
352 & Gisela & 24 & $-$21 & 206 & $-$28 & 7.48008 & 6 & 4 & 134 & 140 &  \\
371 & Bohemia & 93 & 49 & 256 & 43 & 10.73965 & 30 & 4 & 181 & 79 &  \\
390 & Alma & 53 & $-$50 & 275 & $-$76 & 3.74117 & 5 & 2 & 142 & 58 &  \\
403 & Cyane & 65 & 35 & 230 & 33 & 12.2700 & 7 & 3 & 186 & 104 &  \\
404 & Arsinoe & 25 & 57 &  &  & 8.88766 & 49 & 9 & 199 & 104 &  \\
406 & Erna & 357 & $-$49 & 161 & $-$60 & 8.79079 & 8 & 1 & 134 & 93 &  \\
441 & Bathilde & 285 & 55 & 122 & 43 & 10.44313 & 32 & 7 & 158 & 112 &  \\
507 & Laodica & 102 & $-$55 & 312 & $-$49 & 4.70657 &  &  & 162 & 103 &  \\
509 & Iolanda & 245 & 65 & 98 & 38 & 12.2907 & 4 & 2 & 178 & 85 &  \\
512 & Taurinensis & 324 & 45 &  &  & 5.58203 & 11 & 2 & 124 & 111 &  \\
519 & Sylvania & 106 & 9 & 286 & $-$13 & 17.9647 & 5 & 2 & 147 & 76 &  \\
528 & Rezia & 176 & $-$59 & 46 & $-$66 & 7.33797 & 6 & 2 & 151 & 77 &  \\
531 & Zerlina & 78 & $-$84 &  &  & 16.7073 & 28 & 3 & 48 & 52 &  \\
543 & Charlotte & 333 & 59 & 172 & 49 & 10.7184 & 4 & 1 & 138 & 98 &  \\
572 & Rebekka & 1 & 54 & 158 & 39 & 5.65009 & 5 & 2 & 155 & 63 &  \\
578 & Happelia & 339 & 62 &  &  & 10.06450 & 20 & 4 & 183 & 80 &  \\
600 & Musa & 0 & $-$74 & 208 & $-$46 & 5.88638 & 23 & 7 & 96 & 132 &  \\
669 & Kypria & 31 & 40 & 189 & 49 & 14.2789 & 5 & 1 & 142 & 126 &  \\
708 & Raphaela & 37 & 27 & 217 & 22 & 20.8894 & 5 & 1 & 140 & 95 &  \\
725 & Amanda & 145 & $-$63 & 320 & $-$70 & 3.74311 & 18 & 7 & 70 & 77 &  \\
731 & Sorga & 83 & 40 & 275 & 21 & 8.18633 & 7 & 2 & 131 & 136 &  \\
732 & Tjilaki & 160 & 23 & 353 & 24 & 12.3411 & 3 & 1 & 140 & 153 &  \\
787 & Moskva & 331 & 59 & 126 & 27 & 6.05581 & 15 & 4 & 160 & 92 &  \\
792 & Metcalfia & 88 & $-$14 & 274 & $-$13 & 9.17821 & 9 & 3 & 164 & 56 &  \\
803 & Picka & 218 & 34 & 53 & 41 & 5.07478 &  &  & 154 & 50 &  \\
807 & Ceraskia & 325 & 23 & 132 & 26 & 7.37390 & 2 & 1 & 132 & 111 &  \\
812 & Adele & 301 & 44 & 154 & 69 & 5.85746 &  &  & 65 & 119 & \\
816 & Juliana & 124 & $-$8 & 304 & 10 & 10.5627 & 11 & 2 & 158 & 107 &  \\
819 & Barnardiana & 169 & 46 & 334 & 47 & 66.698 &  &  & 121 & 86 &  \\
852 & Wladilena & 181 & $-$48 & 46 & $-$53 & 4.613301 & 30 & 8 & 138 & 101 &  \\
857 & Glasenappia & 227 & 48 & 38 & 34 & 8.20757 & 4 & 2 & 140 & 116 &  \\
867 & Kovacia & 200 & $-$44 & 38 & $-$50 & 8.67807 &  &  & 78 & 76 &  \\
874 & Rotraut & 201 & $-$41 & 2 & $-$36 & 14.3007 & 3 & 1 & 129 & 68 &  \\
875 & Nymphe & 42 & 31 & 196 & 42 & 12.6213 & 6 & 1 & 94 & 100 &  \\
900 & Rosalinde & 276 & 70 & 90 & 39 & 16.6868 & 3 & 2 & 125 & 170 &  \\
920 & Rogeria & 238 & $-$15 & 47 & $-$35 & 12.5749 &  &  & 137 & 79 &  \\
958 & Asplinda & 41 & 48 & 226 & 35 & 25.3050 & 2 & 1 & 98 & 68 &  \\
994 & Otthild & 183 & $-$50 & 41 & -39 & 5.94819 & 26 & 5 & 140 & 125 &  \\ 
1040 & Klumpkea & 172 & 48 &  &  & 56.588 &  &  & 114 & 88 &  \\
1056 & Azalea & 252 & 51 & 64 & 41 & 15.0276 & 3 & 1 & 122 & 112 &  \\
1089 & Tama & 193 & 32 & 9 & 28 & 16.4461 & 90 & 7 & 108 & 79 &  \\
1111 & Reinmuthia & 356 & 68 & 153 & 78 & 4.007347 & 13 & 3 & 137 & 65 &  \\
1126 & Otero & 44 & 75 & 240 & 56 & 3.64800 & 2 & 1 & 101 & 110 &  \\
1130 & Skuld & 24 & 36 & 200 & 35 & 4.80764 & 14 & 1 & 92 & 106 &  \\
1188 & Gothlandia & 334 & $-$84 &  &  & 3.491820 & 36 & 5 & 134 & 91 &  \\
1241 & Dysona & 125 & $-$68 &  &  & 8.60738 & 7 & 1 & 156 & 64 &  \\
1249 & Rutherfordia & 32 & 74 & 197 & 65 & 18.2183 & 6 & 2 & 187 & 75 &  \\
1317 & Silvretta & 45 & $-$57 & 161 & $-$46 & 7.06797 & 13 & 3 & 120 & 69 &  \\
1386 & Storeria & 227 & $-$67 & 297 & $-$67 & 8.67795 & 10 & 1 & 33 & 78 &  \\
1389 & Onnie & 183 & $-$75 & 0 & $-$79 & 23.0447 & 2 & 1 & 90 & 97 &  \\
1393 & Sofala & 319 & 28 & 134 & 41 & 16.5931 &  &  & 69 & 91 &  \\
1401 & Lavonne & 204 & 23 & 27 & 44 & 3.93261 & 3 & 1 & 109 & 88 &  \\
1432 & Ethiopia & 41 & 44 & 225 & 54 & 9.84425 & 11 & 1 & 88 & 101 &  \\
1436 & Salonta & 223 & 18 & 57 & 35 & 8.86985 & 10 & 2 & 132 & 90 &  \\
1450 & Raimonda & 231 & $-$56 & 71 & $-$60 & 12.6344 &  &  & 74 & 116 &  \\
1472 & Muonio & 249 & 61 & 42 & 62 & 8.70543 & 6 & 1 & 99 & 93 &  \\
1490 & Limpopo & 319 & 22 & 142 & 2 & 6.65164 & 5 & 1 & 103 & 107 &  \\
1495 & Helsinki & 355 & $-$39 &  &  & 5.33131 & 13 & 2 & 62 & 109 &  \\
1518 & Rovaniemi & 62 & 60 & 265 & 45 & 5.25047 & 2 & 1 & 100 & 73 &  \\
1528 & Conrada & 250 & $-$51 & 93 & $-$66 & 6.32154 & 2 & 1 & 93 & 126 &  \\
1554 & Yugoslavia & 281 & $-$34 & 78 & $-$64 & 3.88766 & 3 & 1 & 75 & 75 &  \\
1559 & Kustaanheimo & 275 & 29 & 94 & 33 & 4.30435 &  &  & 53 & 82 &  \\
1572 & Posnania & 205 & $-$82 & 85 & $-$63 & 8.04945 & 46 & 7 & 141 & 83 &  \\
1607 & Mavis & 0 & 59 & 222 & 70 & 6.14775 & 4 & 1 & 141 & 179 &  \\
1630 & Milet & 304 & 34 & 121 & 40 & 32.485 & 3 & 1 & 72 & 92 &  \\
1634 & Ndola & 261 & 45 & 66 & 34 & 64.255 & 7 & 1 & 71 & 110 &  \\
1704 & Wachmann & 267 & 41 & 90 & 40 & 3.31391 &  &  & 54 & 135 &  \\
1715 & Salli & 95 & $-$24 & 254 & $-$48 & 11.08867 & 2 & 1 & 84 & 97 &  \\
1719 & Jens & 286 & $-$88 & 55 & $-$42 & 5.87016 & 4 & 2 & 78 & 53 &  \\
1785 & Wurm & 11 & 57 & 192 & 47 & 3.26934 & 2 & 1 & 43 & 115 &  \\
1837 & Osita & 167 & $-$64 & 352 & $-$54 & 3.81879 &  &  & 82 & 62 &  \\
1905 & Ambartsumian & 52 & $-$64 & 241 & $-$68 & 92.153 &  &  & 50 & 101 &  \\
1927 & Suvanto & 74 & 73 & 278 & 23 & 8.16154 & 4 & 1 & 64 & 119 &  \\
1933 & Tinchen & 113 & 26 & 309 & 36 & 3.67062 &  &  & 72 & 103 &  \\
1950 & Wempe & 90 & $-$41 & 258 & $-$45 & 16.7953 & 1 & 1 & 96 & 46 &  \\
1963 & Bezovec & 219 & 7 &  &  & 18.1655 & 12 & 2 & 103 & 40 &  \\
1996 & Adams & 107 & 55 &  &  & 3.31114 &  &  & 82 & 120 &  \\
2002 & Euler & 30 & 44 & 188 & 47 & 5.99264 & 7 & 2 &  & 85 &  \\
2094 & Magnitka & 107 & 57 & 272 & 48 & 6.11219 &  &  & 25 & 84 &  \\
2510 & Shandong & 256 & 27 & 71 & 27 & 5.94639 & 4 & 1 &  & 132 &  \\
2606 & Odessa & 25 & $-$81 & 283 & $-$88 & 8.2444 & 3 & 1 & 25 & 129 &  \\
2709 & Sagan & 302 & $-$14 & 124 & $-$35 & 5.25636 & 6 & 2 &  & 160 &  \\
2839 & Annette & 341 & $-$49 & 154 & $-$36 & 10.4609 & 8 & 1 &  & 99 &  \\
2957 & Tatsuo & 81 & 45 & 248 & 32 & 6.82043 & 13 & 1 & 33 & 102 &  \\
2991 & Bilbo & 277 & 54 & 90 & 51 & 4.06175 & 3 & 1 &  & 97 &  \\
3722 & Urata & 260 & $-$22 & 77 & $-$9 & 5.5671 & 10 & 3 &  & 70 &  \\
4954 & Eric & 86 & $-$54 &  &  & 12.05207 & 7 & 2 &  & 68 &  \\
5281 & Lindstrom & 238 & $-$72 & 84 & $-$81 & 9.2511 & 2 & 1 &  & 76 &  \\
7517 & 1989 AD & 314 & $-$60 & 123 & $-$51 & 9.7094 & 4 & 1 &  & 81 &  \\
8132 & Vitginzburg & 33 & $-$66 & 193 & $-$48 & 7.27529 & 3 & 1 &  & 100 &  \\
8359 & 1989 WD & 121 & $-$68 & 274 & $-$68 & 2.89103 & 6 & 1 &  & 105 &  \\
10772 & 1990 YM & 16 & 46 &  &  & 68.82 & 5 & 1 &  & 73 &  \\
31383 & 1998 XJ$_{94}$ & 110 & $-$74 & 279 & $-$63 & 4.16818 & 4 & 1 &  & 71 &  \\
52820 & 1998 RS$_2$ & 228 & $-$57 & 58 & $-$48 & 2.13412 & 1 & 1 &  & 45 &  \\
57394 & 2001 RD$_{84}$ & 65 & 68 & 241 & 59 & 6.7199 & 4 & 1 &  & 47 &  \\
\hline
\end{longtable}
\tablefoot{
For each asteroid, the table gives the ecliptic coordinates $\lambda_1$ and $\beta_1$ of the pole solution with the lowest chi-square, the corresponding mirror solution $\lambda_2$ and $\beta_2$, the sidereal rotational period $P$, the number of dense lightcurves $N_{\mathrm{lc}}$ observed during $N_{\mathrm{app}}$ apparitions, and the number of sparse data points for the corresponding observatory: $N_{\mathrm{689}}$, $N_{\mathrm{703}}$ and $N_{\mathrm{950}}$. The uncertainty of the sidereal rotational period corresponds to the last decimal place of $P$ and of the pole direction to 10--20$^{\circ}$.
}
}

\begin{table*}
\caption{\label{tab:CSS}List of new asteroid models derived from the Catalina Sky Survey data alone.}
\begin{tabular}{r@{\,\,\,}l rrrr D{.}{.}{6} r D{.}{.}{6} c}\hline 
\multicolumn{2}{c} {Asteroid} & \multicolumn{1}{c} {$\lambda_1$} & \multicolumn{1}{c} {$\beta_1$} & \multicolumn{1}{c} {$\lambda_2$} & \multicolumn{1}{c} {$\beta_2$} & \multicolumn{1}{c} {$P$} & $N_{\mathrm{703}}$ & \multicolumn{1}{c} {$P_{\mathrm{publ}}$} & Period reference\\
\multicolumn{2}{l} { } & [deg] & [deg] & [deg] & [deg] & \multicolumn{1}{c} {[hours]} &  & \multicolumn{1}{c} {[hours]} &  \\
\hline\hline
2112 & Ulyanov & 156 & 48 & 334 & 65 & 3.04071 & 118 & 3.000 & \citet{Maleszewski2004} \\
2384 & Schulhof & 196 & $-$60 & 45 & $-$42 & 3.29367 & 121 & 3.294 & \citet{Ditteon2002} \\
2617 & Jiangxi & 224 & 76 & 1 & 54 & 11.7730 & 124 & 11.79 & \citet{Carbo2009} \\
3170 & Dzhanibekov & 217 & 60 & 21 & 64 & 6.07168 & 105 & 6.0724 & \citet{Molnar2008} \\
4507 & 1990 FV & 143 & 55 & 323 & 49 & 6.57933 & 84 & 6.58 & \citet{Yoshida2005} \\
5647 & 1990 TZ & 253 & 77 & 119 & $-$19 & 6.13867 & 87 & 6.144 & \citet{Bembrick2003} \\
10826 & 1993 SK$_{16}$ & 260 & $-$56 & 60 & $-$34 & 13.8327 & 90 & 13.835 & \citet{Galad2008} \\
19848 & Yeungchuchiu & 190 & $-$68 &   &   & 3.45104 & 104 & 3.450 & \citet{Yeung2006} \\
3097 & Tacitus & 229 & 71 & 72 & 62 & 8.7759 & 99 &  &  \\
4611 & Vulkaneifel & 5 & $-$86 & 197 & $-$50 & 3.75635 & 148 &  &  \\
5461 & Autumn & 249 & $-$26 & 79 & $-$43 & 20.0929 & 106 &  &  \\
5625 & 1991 AO$_{2}$ & 265 & $-$52 & 97 & $-$78 & 6.67411 & 110 &  &  \\
5960 & Wakkanai & 226 & $-$69 & 69 & $-$61 & 4.96286 & 102 &  &  \\
7201 & Kuritariku & 22 & 67 & 249 & 64 & 48.849 & 103 &  &  \\
7632 & Stanislav & 234 & $-$50 & 46 & $-$45 & 5.29073 & 99 &  &  \\
7905 & Juzoitami & 105 & $-$76 & 226 & $-$55 & 2.72744 & 118 &  &  \\
13002 & 1982 BJ$_{13}$ & 58 & $-$50 & 245 & $-$57 & 3.13844 & 110 &  &  \\
16009 & 1999 CM$_{8}$ & 283 & 44 &  &  & 8.3476 & 124 &  &  \\
16847 & Sanpoloamosciano & 91 & $-$24 &  &  & 8.1845 & 114 &  &  \\
26792 & 1975 LY & 226 & 68 &  &  & 79.15 & 140 &  &  \\
\hline
\end{tabular}
\tablefoot{
For each asteroid, the table gives the ecliptic coordinates $\lambda_1$ and $\beta_1$ of the pole solution, the corresponding mirror solution $\lambda_2$ and $\beta_2$, the sidereal rotational period $P$, the number of sparse data points from the CSS $N_{\mathrm{703}}$, and the previously published period value $P_{\mathrm{publ}}$ with the reference. The uncertainty of the sidereal rotational period corresponds to the last decimal place of $P$ and of  the pole direction to 20--40$^{\circ}$.}
\end{table*}

\onecolumn
\scriptsize{
\longtab{3}{
\begin{longtable}{r@{\,\,\,}l lcc}
\caption{\label{tab:references}Observations used for the successful model determinations that are not included in the UAPC.}\\
\hline
 \multicolumn{2}{c} {Asteroid} & Date & Observer & Observatory (MPC code) \\ \hline\hline

\endfirsthead
\caption{continued.}\\

\hline
 \multicolumn{2}{c} {Asteroid} & Date & Observer & Observatory (MPC code) \\
\hline\hline
\endhead
\hline
\endfoot
11 & Parthenope & 2008 5 -- 2008 9 & Warner & Palmer Divide Observatory (716) \\
   &            & 2008 7 -- 2008 7 & Pilcher\tablefootmark{b} & Organ Mesa Observatory (G50) \\
   &            & 2009 11 -- 2010 1 & \citet{Pilcher2010c} &  \\
   &            & 2011 2 -- 2011 5 & \citet{Pilcher2011c} &  \\
   &            & 2011 3 -- 2011 3 & Audejean & Observatoire de Chinon, France (B92) \\
   &            & 2011 4 -- 2011 4 & Naves & Observatorio Montcabre (213) \\
25 & Phocaea    & 2006 10 -- 2006 10 & Buchheim & Altimira Observatory, USA (G76) \\
   &            & 2006 10 21.9 & Strajnic, Grangeon, Coupier, Godon, Roche & Haute-Provence Observatory, France (511) \\
   &            &              & Danavaro, Dalmas, Bayol, Behrend &  \\
   &            & 2008 1 -- 2009 4 & \citet{Pilcher2009a} &  \\
   &            & 2010 9 -- 2010 12 & \citet{Pilcher2011b} &  \\
72 & Feronia    & 2004 3 -- 2004 4 & Bernasconi & Les Engarouines Observatory, France (A14) \\
   &            & 2005 7 -- 2005 8 & Bernasconi & Les Engarouines Observatory, France (A14) \\
   &            & 2007 1 20.9 & Coliac & Observatoire Farigourette, France \\
   &            & 2011 3 -- 2011 4 & Marciniak & Borowiec, Poland (187) \\
   &            & 2011 5 9.9 & Hirsch & Borowiec, Poland (187) \\
147 & Protogeneia & 2004 11 -- 2004 12 & \citet{Buchheim2005} &  \\
    &             & 2005 1 4.9 & Roy & Blauvac Observatory, France (627) \\
    &             & 2005 1 --2005 1 & Bernasconi & Les Engarouines Observatory, France (A14) \\ 
    &             & 2008 5 29.7 & Higgins\tablefootmark{a} & Hunters Hill Observatory, Ngunnawal (E14) \\
149 & Medusa    & 2010 10 -- 2010 11 & \citet{Pilcher2011b} &  \\
    &           & 2010 11 -- 2010 12 & Martin & Tzec Maun Observatory, Mayhill (H10) \\
157 & Dejanira  & 2005 3 -- 2005 3 & Poncy & Le Cr\'es, France (177) \\
    &           & 2005 4 -- 2005 5 & \citet{Warner2005b} &  \\
   &            & 2008 12 -- 2009 2 & \citet{Pilcher2009c} &  \\
166 & Rhodope   & 2010 12 -- 2011 1 & Conjat & Cabris, France \\
178 & Belisana  & 2007 4-2007 7 & \citet{Oey2008} &  \\
   &            & 2008 9 -- 2008 10 & \citet{Pilcher2009d} &  \\
183 & Istria    & 2004 2 14.1 & Bernasconi & Les Engarouines Observatory, France (A14) \\
193 & Ambrosia  & 2009 4 -- 2009 4 & \citet{Warner2009b} &  \\
    &           & 1999 10 15.0 & Hirsch & Borowiec, Poland (187) \\
    &           & 2005 4 -- 2005 4 & Kaminski & Borowiec, Poland (187) \\
    &           & 2005 4 3.9 & Marciniak & Borowiec, Poland (187) \\
    &           & 2005 4 -- 2005 4 & Hirsch & Borowiec, Poland (187) \\
    &           & 2009 3 -- 2009 3 & Audejean & Observatoire de Chinon, France (B92) \\
    &           & 2009 4 -- 2009 5 & Hirsch & Borowiec, Poland (187) \\
    &           & 2009 4 29.9 & Kaminski & Borowiec, Poland (187) \\
    &           & 2010 4 19.1 & Borczyk & SAAO, Sutherland, South Africa \\
199 & Byblis    & 2003 3 -- 2003 4 & Casulli & Vallemare di Bordona, Italy (A55) \\
    &           & 2003 5 -- 2003 5 & Bernasconi & Les Engarouines Observatory, France (A14) \\
    &           & 2005 10 -- 2005 10 & Roy & Blauvac Observatory, France (627) \\
    &           & 2005 10 -- 2005 10 & Casulli & Vallemare di Bordona, Italy (A55) \\
    &           & 2005 11 -- 2005 11 & Stoss, Nomen, Sanchez, Behrend & OAM - Mallorca (620) \\
    &           & 2005 11 20.9 & Farroni &  \\
    &           & 2006 12 -- 2006 12 & Roy & Blauvac Observatory, France (627) \\
    &           & 2008 2 9.1 & Manzini & Stazione Astronomica di Sozzago, Italy (A12) \\
    &           & 2011 9 24.1 & Sobkowiak & Borowiec, Poland (187) \\
    &           & 2011 11 -- 2011 11 & Marciniak & Borowiec, Poland (187) \\
220 & Stephania & 2004 10 -- 2004 10 & Koff & Antelope Hills Observatory, Bennett (H09) \\
222 & Lucia     & 1999 4 18.2 & Warner & Palmer Divide Observatory (716) \\
    &           & 2008 12 -- 2008 12 & \citet{Stephens2009} &  \\
   &            & 2010 4 -- 2010 5 & Audejean & Observatoire de Chinon, France (B92) \\
   &            & 2010 4 -- 2010 4 & Bosch & Collonges Observatory, France (178) \\
242 & Kriemhild & 2004 7 -- 2004 7 & Bosch & Collonges Observatory, France (178) \\
    &           & 2004 8 -- 2004 8 & \citet{Warner2005c} &  \\
    &           & 2004 9 -- 2004 9 & Rinner & Ottmarsheim Observatory, France (224) \\
    &           & 2005 11 7.9 & Roy & Blauvac Observatory, France (627) \\
    &           & 2007 1 -- 2007 1 & \citet{Bembrick2007a} &  \\
    &           & 2009 8 -- 2009 8 & Audejean & Observatoire de Chinon, France (B92) \\
    &           & 2010 8 -- 2011 3 & Marciniak & Borowiec, Poland (187) \\
    &           & 2010 10 10.1 & T.~Micha{\l}owski & Borowiec, Poland (187) \\
    &           & 2011 11 -- 2012 1 & Marciniak & Borowiec, Poland (187) \\
    &           & 2011 11 13.1 & Sobkowiak & Borowiec, Poland (187) \\
257 & Silesia   & 2004 12 -- 2004 12 & Casulli, Behrend & Vallemare di Bordona, Italy (A55) \\
   &            & 2004 12 -- 2005 1 & Roy & Blauvac Observatory, France (627) \\
   &            & 2005 1 31.1 & Starkey & DeKalb Observatory, USA (H63) \\
   &            & 2005 12 1.1 & Strajnic, Paulo, Wagrez, Jade, & Haute-Provence Observatory, France (511) \\
   &            &  & Rocca, Del Freo, Behrend &  \\
   &            & 2005 12 --2006 1 & Roy & Blauvac Observatory, France (627) \\
   &            & 2005 12 -- 2005 12 & Antonini & Observatoire de B\'edoin, France (132) \\
260 & Huberta   & 2005 3 -- 2005 3 & Roy & Blauvac Observatory, France (627) \\
   &            & 2007 7 -- 2007 8 & Roy & Blauvac Observatory, France (627) \\
272 & Antonia   & 2007 12 -- 2008 1 & \citet{Pilcher2008a} &  \\
    &           & 2011 10 -- 2011 10 & S.~Fauvaud, M.~Fauvaud & Observatoire du Bois de Bardon, France \\
281 & Lucretia  & 2011 10 -- 2011 10 & S.Fauvaud, M.~Fauvaud & Observatoire du Bois de Bardon, France \\
290 & Bruna     & 2008 3 -- 2008 4 & \citet{Pilcher2009e} &  \\
297 & Caecilia  & 2004 11 -- 2004 12 & Roy & Blauvac Observatory, France (627) \\
    &           & 2006 1 -- 2006 1 & Manzini & Stazione Astronomica di Sozzago, Italy (A12) \\
    &           & 2006 1 11.0 & Antonini & Observatoire de B\'edoin, France (132) \\
    &           & 2006 1 13.1 & Roy & Blauvac Observatory, France (627) \\
    &           & 2009 12 11.8 & Salom, Esteban & Caimari (B81) \\
    &           & 2011 2 -- 2011 3 & Marciniak & Borowiec, Poland (187) \\
    &           & 2012 1 30.2 & Marciniak & Borowiec, Poland (187) \\
    &           & 2012 1 31.2 & Polinska & Borowiec, Poland (187) \\
    &           & 2012 2 -- 2012 3 & Hirsch & Borowiec, Poland (187) \\
345 & Tercidina & 2002 9 -- 2002 10 & Barbotin & Villefagnan Observatory, France \\
    &           & 2002 9 -- 2002 12 & Bernasconi & Les Engarouines Observatory, France (A14) \\
    &           & 2002 9 -- 2002 10 & Rinner & Ottmarsheim Observatory, France (224) \\
    &           & 2002 9 -- 2002 9 & Starkey, Bernasconi & Les Engarouines Observatory, France (A14) \\
    &           & 2002 9 -- 2002 9 & Waelchli, Revaz & F.-X.~Bagnoud Observatory, Switzerland (175) \\
    &           & 2002 10 1.1 & Michelet &  \\
    &           & 2002 10 5.2 & Barbotin & Villefagnan Observatory, France \\
    &           & 2002 11 22.9 & Bosch & Collonges Observatory, France (178) \\
    &           & 2002 11-2002 12 & Starkey & DeKalb Observatory, USA (H63) \\
    &           & 2004 4 -- 2004 5 & Bernasconi & Les Engarouines Observatory, France (A14) \\
    &           & 2004 4 -- 2004 5 & Roy & Blauvac Observatory, France (627) \\
    &           & 2005 8 -- 2005 8 & Bernasconi & Les Engarouines Observatory, France (A14) \\
    &           & 2005 8 27.0 & Stoss, Nomen, Sanchez, Behrend & OAM - Mallorca (620) \\
    &           & 2005 9 8.0 & Farroni &  \\
    &           & 2008 7 5.0 & Tr\'egon, Leroy & Pic du Midi Observatory (586) \\
    &           & 2009 8 -- 2009 10 & Naves & Observatorio Montcabre (213) \\
    &           & 2011 4 22.9 & Sobkowiak & Borowiec, Poland (187) \\
352 & Gisela    & 2002 10 8.7 & Droege &  \\
   &            & 2004 2 13.1 & Bernasconi, Klotz, Behrend & Haute-Provence Observatory, France (511) \\
   &            & 2005 7 -- 2005 8 & Bernasconi & Les Engarouines Observatory, France (A14) \\
371 & Bohemia   & 2001 6 -- 2004 3 & \citet{Buchheim2004a} &  \\
    &           & 2006 9 2.0 & Bernasconi & Les Engarouines Observatory, France (A14) \\
    &           & 2011 8 -- 2011 11 & Marciniak & Borowiec, Poland (187) \\
    &           & 2011 11 2.9 & W.~Og\l oza & Mnt. Suhora, Poland \\
    &           & 2011 11 30.9 & Santana-Ros  & Borowiec, Poland (187) \\
390 & Alma      & 2004 8 -- 2004 8 & \citet{Stephens2005c} &  \\
   &            & 2008 8 -- 2008 10 & Roy & Blauvac Observatory, France (627) \\
403 & Cyane     & 2001 12 9.1 & Brunetto & Le Florian, France (139) \\
   &            & 2001 12 -- 2001 12 & Bernasconi & Les Engarouines Observatory, France (A14) \\
   &            & 2001 12 22.2 & Cooney &  \\
   &            & 2005 10 1.0 & Bernasconi & Les Engarouines Observatory, France (A14) \\
   &            & 2007 2 -- 2007 2 & Roy & Blauvac Observatory, France (627) \\
404 & Arsinoe   & 1999 3 -- 1999 4 & Kryszczynska & Borowiec, Poland (187) \\
    &           & 1999 3 19.0 & Hirsch & Borowiec, Poland (187) \\
    &           & 1999 3 20.0 & T.~Micha{\l}owski & Borowiec, Poland (187) \\
    &           & 2001 10 -- 2001 10 & S.~Fauvaud, Heck, Santacana, Wucher & Pic de Ch\^ ateau-Renard Observatory \\
    &           & 2001 11 -- 2001 12 & Bernasconi & Les Engarouines Observatory, France (A14) \\
    &           & 2003 4 -- 2003 4 & Roy & Blauvac Observatory, France (627) \\
    &           & 2005 8 10.1 & Fagas & Borowiec, Poland (187) \\
    &           & 2005 10 -- 2005 10 & Hirsch & Borowiec, Poland (187) \\
    &           & 2005 10 -- 2005 11 & Roy & Blauvac Observatory, France (627) \\
    &           & 2006 11 -- 2007 1 & Fagas & Borowiec, Poland (187) \\
    &           & 2007 1 -- 2007 4 & Marciniak & Borowiec, Poland (187) \\
    &           & 2007 2 17.0 & Hirsch & Borowiec, Poland (187) \\
    &           & 2007 4 -- 2007 4 & Kaminski & Borowiec, Poland (187) \\
    &           & 2007 4 22.0 & Kankiewicz & Kielce, Poland (B02) \\
    &           & 2008 6 -- 2008 6 & Marciniak & SAAO, Sutherland, South Africa \\
    &           & 2009 8 -- 2009 10 & Marciniak & SAAO, Sutherland, South Africa \\
    &           & 2009 9 27.0 & Hirsch & Borowiec, Poland (187) \\
    &           & 2009 10 30.0 & Polinska & Borowiec, Poland (187) \\
    &           & 2009 12 3.0 & Kaminski & Borowiec, Poland (187) \\
    &           & 2010 12 5.0 & Sobkowiak & Borowiec, Poland (187) \\
    &           & 2011 1 -- 2011 5 & Marciniak & Borowiec, Poland (187) \\
    &           & 2011 3 -- 2011 3 & Hirsch & Borowiec, Poland (187) \\
406 & Erna      & 2005 9 -- 2005 10 & Casulli & Vallemare di Bordona, Italy (A55) \\
   &            & 2005 11 -- 2005 11 & Crippa, Manzini & Stazione Astronomica di Sozzago, Italy (A12) \\
   &            & 2005 11 -- 2005 11 & Poncy & Le Cr\'es, France (177) \\
441 & Bathilde  & 2003 1 -- 2003 1 & Roy & Blauvac Observatory, France (627) \\
    &           & 2003 2 -- 2003 2 & Bernasconi & Les Engarouines Observatory, France (A14) \\
    &           & 2003 2 -- 2003 3 & Vagnozzi, Cristofanelli, Paiella & Santa Lucia Stroncone (589) \\
    &           & 2005 7 -- 2005 8 & Bernasconi & Les Engarouines Observatory, France (A14) \\
    &           & 2006 12 11.9 & Poncy & Le Cr\'es, France (177) \\
    &           & 2010 9 -- 2010 10 & Marciniak & Borowiec, Poland (187) \\
    &           & 2010 10 4.8 & Kaminski & Borowiec, Poland (187) \\
    &           & 2010 10 9.9 & T.~Micha{\l}owski & Borowiec, Poland (187) \\
    &           & 2011 10 14.0 & Sobkowiak & Borowiec, Poland (187) \\
    &           & 2011 10 -- 2011 11 & Marciniak & Borowiec, Poland (187) \\
507 & Laodica   & 2001 8 -- 2001 8 & Charbonnel & Durtal (949) \\
    &           & 2001 8 -- 2001 9 & Leyrat &  \\
509 & Iolanda   & 1996 10 -- 1996 10 & \citet{Lopez2000} & \\
    &           & 2000 6 8.3 & \citet{Koff2000} &  \\
512 & Taurinensis & 2004 12 -- 2005 1 & Poncy & Le Cr\'es, France (177) \\
    &           & 2005 1 5.0 & Correia & Haute-Provence Observatory, France (511) \\
528 & Rezia     & 2011 3 -- 2011 3 & Mottola &  \\
531 & Zerlina   & 2002 6 2.9 & Christophe &  \\
    &           & 2007 9 -- 2007 10 & \citet{Brinsfield2008} &  \\
    &           & 2011 3 -- 2011 6 & \citet{Pilcher2011d}  &  \\
543 & Charlotte & 2006 11 -- 2006 12 & Poncy & Le Cr\'es, France (177) \\
572 & Rebekka   & 2007 2 -- 2007 2 & \citet{Warner2007} &  \\
    &           & 2009 8 -- 2009 8 & Audejean & Observatoire de Chinon, France (B92) \\
578 & Happelia  & 2006 12 -- 2006 12 & Leroy & Uranoscope, France (A07) \\
    &           & 2008 4 -- 2008 4 & \citet{Warner2008c} &  \\
    &           & 2010 11 -- 2010 12 & Antonini & Observatoire de B\'edoin, France (132) \\
    &           & 2012 2 -- 2012 4 & Mottola, Hellmich &  \\
600 & Musa      & 2001 4 6.0 & Hirsch & Borowiec, Poland (187) \\
    &           & 2001 4 29.0 & Colas & Pic du Midi Observatory (586) \\
    &           & 2005 2 -- 2005 3 & Bernasconi & Les Engarouines Observatory, France (A14) \\
    &           & 2005 3 -- 2005 4 & Hirsch & Borowiec, Poland (187) \\
    &           & 2005 4 1.0 & Marciniak & Borowiec, Poland (187) \\
    &           & 2007 10 -- 2007 10 & S.~Fauvaud, Santacana, M.~Fauvaud & Pic du Midi Observatory (586) \\
    &           & 2009 3 25.8 & Kaminski & Borowiec, Poland (187) \\
    &           & 2009 3 30.9 & Marciniak & Borowiec, Poland (187) \\
    &           & 2010 4 -- 2010 6 & Marciniak & Borowiec, Poland (187) \\
    &           & 2011 11 -- 2011 11 & Marciniak & Borowiec, Poland (187) \\
    &           & 2011 11 29.8 & Hirsch & Borowiec, Poland (187) \\
669 & Kypria    & 2006 3 -- 2006 4 & Bernasconi & Les Engarouines Observatory, France (A14) \\
708 & Raphaela  & 2007 2 -- 2007 2 & \citet{Warner2007} &  \\
725 & Amanda    & 2002 12 12.8 & Marciniak & Borowiec, Poland (187) \\
    &           & 2002 12 31.8 & T.~Micha{\l}owski & Borowiec, Poland (187) \\
    &           & 2006 10 -- 2006 10 & S.~Fauvaud, Santacana, Sareyan, Wucher & Pic de Ch\^ ateau-Renard Observatory \\
    &           & 2006 10 30.1 & Hirsch & Borowiec, Poland (187) \\
    &           & 2009 8 -- 2009 8 & Marciniak & SAAO, Sutherland, South Africa \\
    &           & 2010 10 -- 2010 10 & Audejean & Observatoire de Chinon, France (B92) \\
    &           & 2010 10 31.0 & Marciniak & Borowiec, Poland (187) \\
    &           & 2012 3 3.1 & Marciniak & Borowiec, Poland (187) \\
    &           & 2012 3 -- 2012 3 & Hirsch & Borowiec, Poland (187) \\
    &           & 2012 4 10.1 & Oszkiewicz, Geier & NOT, La Palma, Canary Islands \\
731 & Sorga     & 2005 4 -- 2005 4 & \citet{Warner2005b} &  \\
    &           & 2009 2 -- 2009 2 & \citet{Warner2009c} &  \\
732 & Tjilaki   & 2004 3 -- 2004 4 & Bernasconi & Les Engarouines Observatory, France (A14) \\
787 & Moskva    & 1999 5 -- 1999 5 & \citet{Warner2011b} &  \\
    &           & 2003 4 -- 2003 5 & Husarik, Behrend & Skalnate Pleso, Slovakia (056) \\
    &           & 2003 5 -- 2003 5 & Bernasconi & Les Engarouines Observatory, France (A14) \\
    &           & 2004 8 -- 2004 8 & Bernasconi & Les Engarouines Observatory, France (A14) \\
    &           & 2011 5 -- 2011 5 & Audejean & Observatoire de Chinon, France (B92) \\
    &           & 2011 5 -- 2011 5 & Morelle & Observatoire Farigourette, France \\
792 & Metcalfa  & 2010 7 -- 2010 8 & Roy & Blauvac Observatory, France (627) \\
803 & Picka     & 2006 12 10.8 & Bosch & Collonges Observatory, France (178) \\
    &           & 2007 4 -- 2007 4 & Antonini & Observatoire de B\'edoin, France (132) \\
    &           & 2010 11 -- 2010 11 & Antonini & Observatoire de B\'edoin, France (132) \\
812 & Adele     & 2002 10 -- 2002 10 & Roy & Blauvac Observatory, France (627) \\
816 & Juliana   & 2005 4 -- 2005 4 & \citet{Stephens2005a} &  \\
    &           & 2005 5 -- 2005 6 & Conjat & Cabris, France \\
    &           & 2010 3 -- 2010 3 & Conjat & Cabris, France \\
852 & Wladilena & 2003 2 23.2 & J.~Micha{\l}owski & Borowiec, Poland (187) \\
    &           & 2003 2 24.2 & Marciniak & Borowiec, Poland (187) \\
    &           & 2003 2 26.2 & T.~Micha{\l}owski & Borowiec, Poland (187) \\
    &           & 2007 5 -- 2007 5 & Marciniak & SAAO, Sutherland, South Africa \\
    &           & 2008 8 22.2 & M.~J.~Micha{\l}owski & NOT, La Palma, Canary Islands \\
    &           & 2008 10 -- 2009 1 & Kaminski & Borowiec, Poland (187) \\
    &           & 2008 9 -- 2008 10 & Marciniak & Borowiec, Poland (187) \\
    &           & 2008 12 -- 2009 1 & Sobkowiak & Borowiec, Poland (187) \\
    &           & 2010 2 -- 2010 3 & Antonini & Observatoire de B\'edoin, France (132) \\
    &           & 2010 3 -- 2010 5 & Marciniak & Borowiec, Poland (187) \\
    &           & 2010 3 -- 2010 3 & \citet{Polishook2012b}\tablefootmark{c} & Wise Observatory, Mitzpeh Ramon (097) \\
    &           & 2010 3 -- 2010 4 & Sobkowiak & Borowiec, Poland (187) \\
857 & Glasenapia & 2006 12 23.0 & Poncy & Le Cr\'es, France (177) \\
867 & Kovacia   & 2006 11 22.8 & Crippa, Manzini & Stazione Astronomica di Sozzago, Italy (A12) \\
    &           & 2008 1 --2008 2 & Roy & Blauvac Observatory, France (627) \\
    &           & 2008 2 8.9 & Casulli & Vallemare di Bordona, Italy (A55) \\
    &           & 2008 2 9.0 & Colas & Pic du Midi Observatory (586) \\
    &           & 2008 2 -- 2008 2 & Manzini & Stazione Astronomica di Sozzago, Italy (A12) \\
    &           & 2008 2 -- 2008 2 & Leroy & Uranoscope, France (A07) \\
    &           & 2008 2 -- 2008 2 & Demeautis & Village-Neuf Observatory, France (138) \\
    &           & 2008 2 -- 2008 3 & Coliac & Observatoire Farigourette, France \\
874 & Rotraut   & 2002 7 -- 2002 7 & Charbonnel & Durtal (949) \\
    &           & 2002 8 16.0 & Rinner & Ottmarsheim Observatory, France (224) \\
875 & Nymphe    & 2003 7 -- 2003 7 & \citet{Warner2011c} &  \\
    &           & 2003 7 -- 2003 7 & Roy & Blauvac Observatory, France (627) \\
900 & Rosalinde & 2007 5 19.0 & Roy & Blauvac Observatory, France (627) \\
994 &  Otthild  & 2001 9 22.0 & Velichko, T.~Micha{\l}owski & Kharkov (101) \\
    &           & 2001 10 -- 2001 10 & J.~Micha{\l}owski & Borowiec, Poland (187) \\
    &           & 2001 10 -- 2001 10 & Conjat & Cabris, France \\
    &           & 2001 11 -- 2001 11 & T.~Micha{\l}owski & Borowiec, Poland (187) \\
    &           & 2001 11 -- 2001 11 & Kwiatkowski & Borowiec, Poland (187) \\
    &           & 2005 8 -- 2005 11 & Stoss, Nomen, Sanchez, Behrend & OAM - Mallorca (620) \\
    &           & 2005 10 1.9 & Bernasconi & Les Engarouines Observatory, France (A14) \\
    &           & 2005 10 -- 2005 10 & Fagas & Borowiec, Poland (187) \\
    &           & 2005 10 19.9 & T.~Micha{\l}owski & Borowiec, Poland (187) \\
    &           & 2007 2 26.9 & S.~Fauvaud, Esseiva, Michelet, & Pic de Ch\^ ateau-Renard Observatory \\
    &           &  & Saguin, Sareyan &  \\
    &           & 2011 3 19.9  & Polinska & Borowiec, Poland (187) \\
    &           & 2011 3 29.8  & Marciniak & Borowiec, Poland (187) \\
1056 & Azalea   & 2004 2 -- 2004 2 & Klotz, Behrend & Haute-Provence Observatory, France (511) \\
1089 & Tama     & 2003 12 -- 2004 3 & Roy & Blauvac Observatory, France (627) \\
     &          & 2003 12 -- 2004 2 & Rinner & Ottmarsheim Observatory, France (224) \\
     &          & 2004 1 -- 2004 1 & Antonini & Observatoire de B\'edoin, France (132) \\
     &          & 2004 1 -- 2004 1 & Sposetti, Behrend & Gnosca Observatory, Switzerland (143) \\
     &          & 2004 1 4.9 & Klotz & Haute-Provence Observatory, France (511) \\
     &          & 2004 1 -- 2004 1 & Lecacheux, Colas & Pic du Midi Observatory (586) \\
     &          & 2004 1 22.8 & Colas & Pic du Midi Observatory (586) \\
     &          & 2004 1 26.9 & Michelsen, Augustesen, Masi & Brorfelde (054) \\
     &          & 2004 1 --2004 1 & Cotrez, Behrend & Saint-H\'el\`ene Observatory, France (J80) \\
     &          & 2004 1 -- 2004 2 & Durkee & Shed of Science Observatory, USA (H39) \\
     &          & 2004 2 7.9 & Bernasconi & Les Engarouines Observatory, France (A14) \\
     &          & 2004 2 9.8 & Coloma & Sabadell (619) \\
     &          & 2004 2 -- 2004 2 & Oksanen & Nyr\"ol\"a Observatory, Finland (174) \\
     &          & 2004 2 11.9 & Itkonen, P\" a\" akk\" onen & Jakokoski Observatory, Finland (A83) \\
     &          & 2004 2 15.0 & Brochard &  \\
     &          & 2004 2 20.9 & Demeautis, Matter & Village-Neuf Observatory, France (138) \\
     &          & 2004 2 24.1 & Barbotin, Cotrez, Cazenave, Laffont & Pic du Midi Observatory (586) \\
     &          & 2005 6 -- 2005 7 & Stoss, Nomen, Sanchez, Behrend & OAM - Mallorca (620) \\
     &          & 2005 7 -- 2005 8 & Teng, Behrend & Observatoire Les Makes, France (181) \\
     &          & 2006 9 -- 2006 12 & Sposetti, Pavic & Gnosca Observatory, Switzerland (143) \\
     &          & 2006 9 -- 2006 12 & \citet{Polishook2012b}\tablefootmark{c} & Wise Observatory, Mitzpeh Ramon (097) \\
     &          & 2006 11 26.9 & Sposetti, Behrend & Gnosca Observatory, Switzerland (143) \\
     &          & 2008 4 5.1 & Klotz, Strajnic & Haute-Provence Observatory, France (511) \\
     &          & 2008 5 -- 2008 5 & Roy & Blauvac Observatory, France (627) \\
     &          & 2008 5 -- 2008 5 & \citet{Polishook2012b}\tablefootmark{c} & Wise Observatory, Mitzpeh Ramon (097) \\
     &          & 2009 10 -- 2009 11 & \citet{Polishook2012b}\tablefootmark{c} & Wise Observatory, Mitzpeh Ramon (097) \\
     &          & 2011 2 -- 2011 3 & Crippa, Manzini & Stazione Astronomica di Sozzago, Italy (A12) \\
1111 & Reinmuthia & 2007 10 -- 2007 11 & Hiromi Hamanowa, Hiroko Hamanowa &  \\
1126 & Otero    & 2008 2 -- 2008 2 & \citet{Stephens2008} &  \\
1130 & Skuld    & 2004 1 22.0 & Colas & Pic du Midi Observatory (586) \\
     &          & 2009 10 -- 2009 11 & \citet{Buchheim2010} &  \\
1188 & Gothlandia & 2006 1 2.9 & Pallares & Sabadell (619) \\
     &          & 2006 1 11.9 & Coloma & Agrupaci\'on Astron\'omica de Sabadell, Spain (A90) \\
     &          & 2006 2 2.9 & Coloma, Garcia & Agrupaci\'on Astron\'omica de Sabadell, Spain (A90) \\
     &          & 2007 5 -- 2007 5 & Antonini & Observatoire de B\'edoin, France (132) \\
     &          & 2008 12 -- 2009 1 & H.~Hamanowa, H.~Hamanowa &  \\
     &          & 2011 8 -- 2011 12 & \citet{Baker2012} &  \\
     &          & 2011 9 -- 2011 9 & S.~Fauvaud, M.~Fauvaud & Observatoire du Bois de Bardon, France \\
1241 & Dysona   & 2002 9 --2002 11 & Bosch & Collonges Observatory, France (178) \\
     &          & 2002 10 2.0 & Brunetto & Le Florian, France (139) \\
     &          & 2006 4 -- 2006 5 & Oey & Leura (E17) \\
1249 & Rutherfordia & 2001 8 -- 2001 8 & Bernasconi & Les Engarouines Observatory, France (A14) \\
     &          & 2008 8 22.0 & Demeautis & Village-Neuf Observatory, France (138) \\
     &          & 2004 7 -- 2004 7 & Roy & Blauvac Observatory, France (627) \\
1317 & Silvretta & 2006 4 -- 2006 4 & Bernasconi & Les Engarouines Observatory, France (A14) \\
     &          & 2009 12 -- 2010 1 & \citet{Ruthroff2010} &  \\
1386 & Storeria & 2004 6 -- 2004 6 & \citet{Warner2004} &  \\
     &          & 2004 7 15.0 & Behrend, Klotz & Haute-Provence Observatory, France (511) \\
     &          & 2004 7 17.0 & Bernasconi & Les Engarouines Observatory, France (A14) \\
     &          & 2004 7 21.0 & Coloma & Agrupaci\'on Astron\'omica de Sabadell, Spain (A90) \\
     &          & 2004 7 28.0 & Roy & Blauvac Observatory, France (627) \\
1401 & Lavonne  & 2008 8 8.3 & Durkee & Shed of Science Observatory, USA (H39) \\
     &          & 2008 9 -- 2008 9 & Antonini & Observatoire de B\'edoin, France (132) \\
1432 & Ethiopia & 2007 7 -- 2007 9 & \citet{Oey2008b} &  \\
1436 & Salonta  & 2007 8 -- 2007 9 & \citet{Warner2008d} &  \\
     &          & 2007 10 -- 2007 10 & Antonini & Observatoire de B\'edoin, France (132) \\
     &          & 2008 11 -- 2008 11 & Antonini & Observatoire de B\'edoin, France (132) \\
     &          & 2008 11 27.8 & Roy & Blauvac Observatory, France (627) \\
1472 & Muonio   & 2008 9 -- 2008 9 & \citet{Stephens2009b} &  \\
     &          & 2008 10 -- 2008 10 & Higgins\tablefootmark{a} & Hunters Hill Observatory, Ngunnawal (E14) \\
1490 & Limpopo  & 2005 8 -- 2005 8 & Bernasconi & Les Engarouines Observatory, France (A14) \\
1495 & Helsinki & 2006 4 -- 2006 5 & \citet{Oey2007} &  \\
     &          & 2006 6 4.0 & Payet, Teng, Leonie, Behrend & Observatoire Les Makes, France (181) \\
     &          & 2006 6 -- 2006 7 & Teng, Behrend & Observatoire Les Makes, France (181) \\
     &          & 2011 9 -- 2011 9 & S.~Fauvaud, M.~Fauvaud & Observatoire du Bois de Bardon, France \\
1518 & Rovaniemi & 2009 1 -- 2009 1 & \citet{Warner2009c} &  \\
     &          & 2009 1 -- 2009 1 & Roy & Blauvac Observatory, France (627) \\
1528 & Conrada  & 2008 5 -- 2008 5 & \citet{Warner2008c}  &  \\
1554 & Yugoslavia & 2007 4 -- 2007 4 & \citet{Higgins2008} &  \\
1559 & Kustaanheimo & 2005 3 -- 2005 3 & Bernasconi & Les Engarouines Observatory, France (A14) \\
1572  & Posnania & 1993 9 -- 1999 11 & \citet{Michalowski2001} & \\
    &           & 2004 9 -- 2004 9 & Roy & Borowiec, Poland (187) \\
    &           & 2010 12 5.1 & Sobkowiak & Borowiec, Poland (187) \\
    &           & 2011 2 -- 2011 2 & Kaminski & Borowiec, Poland (187) \\
    &           & 2011 2 8.8 & Marciniak & Borowiec, Poland (187) \\
     &          & 2012 2 -- 2012 3 & Roy & Blauvac Observatory, France (627) \\
1607 & Mavis    & 2007 9 -- 2007 9 & \citet{Oey2008b} &  \\
1630 & Milet    & 2005 2 -- 2005 2 & Bernasconi & Les Engarouines Observatory, France (A14) \\
1634 & Ndola    & 2006 9 -- 2006 9 & Higgins\tablefootmark{a} & Hunters Hill Observatory, Ngunnawal (E14) \\
1719 & Jens     & 2000 9 -- 2000 9 & \citet{Warner2011a} &  \\
     &          & 2006 1 -- 2006 2 & Bernasconi & Les Engarouines Observatory, France (A14) \\
1785 & Wurm     & 2008 3 -- 2008 3 & \citet{Oey2009} &  \\
1837 & Osita    & 2006 1 -- 2006 3 & Roy & Blauvac Observatory, France (627) \\
1927 & Suvanto  & 2005 2 -- 2005 2 & Bernasconi & Les Engarouines Observatory, France (A14) \\
1933 & Tinchen  & 2005 3 14.0 & Roy & Blauvac Observatory, France (627) \\
1950 & Wempe    & 2006 2 1.9 & Bernasconi & Les Engarouines Observatory, France (A14) \\
1963 & Bezovec  & 2005 1 -- 2005 1 & Bernasconi & Les Engarouines Observatory, France (A14) \\
     &          & 2009 3 -- 2009 3 & Romeuf &  \\
     &          & 2009 4 6.9 & Manzini & Stazione Astronomica di Sozzago, Italy (A12) \\
     &          & 2009 4 -- 2009 4 & Martin & Tzec Maun Observatory, Mayhill (H10) \\
2002 & Euler    & 2006 5 -- 2006 5 & Koff & Antelope Hills Observatory, Bennett (H09) \\
     &          & 2007 10 -- 2007 10 & Higgins\tablefootmark{a} & Hunters Hill Observatory, Ngunnawal (E14) \\
2510 & Shandong & 2006 8 -- 2006 9 & \citet{Higgins2007} &  \\
2606 & Odessa   & 2008 2 -- 2008 2 & \citet{Higgins2008b} &  \\
     &          & 2008 3 3.6 & Oey & Leura (E17) \\
2709 & Sagan    & 2008 3 -- 2008 3 & \citet{Higgins2008b} &  \\
     &          & 2011 1 -- 2011 2 & Oey & Leura (E17) \\
2839 & Annette  & 2005 10 -- 2005 11 & \citet{Buchheim2007} &  \\
     &          & 2005 12 -- 2005 12 & \citet{Warner2006} &  \\
2957 & Tatsuo   & 2005 8 -- 2005 8 & Bernasconi & Les Engarouines Observatory, France (A14) \\
     &          & 2005 8 -- 2005 9 & Poncy & Le Cr\'es, France (177) \\
     &          & 2005 9 -- 2005 9 & \citet{Warner2006b} &  \\
2991 & Bilbo    & 2007 4 -- 2007 4 & Higgins\tablefootmark{a} & Hunters Hill Observatory, Ngunnawal (E14) \\
3722 & Urata    & 2004 12 -- 2004 12 & Antonini & Observatoire de B\'edoin, France (132) \\
     &          & 2006 9 3.0 & Manzini & Stazione Astronomica di Sozzago, Italy (A12) \\
     &          & 2007 8 -- 2007 8 & Roy & Blauvac Observatory, France (627) \\
     &          & 2007 8 -- 2007 8 & Stephens & Goat Mountain Astronomical Research Station (G79) \\
5281 & Lindstrom & 2008 6 -- 2008 6 & Brinsfield & Via Capote Sky Observatory, Thousand Oaks (G69) \\
7517 & 1989 AD  & 2007 11 -- 2007 11 & Stephens & Goat Mountain Astronomical Research Station (G79) \\
8132 & Vitginzburg & 2008 6 -- 2008 6 & \citet{Brinsfield2008b} &  \\
8359 & 1989 WD  & 2009 4 -- 2009 4 & \citet{Higgins2009} &  \\
     &          & 2009 5 -- 2009 5 & \citet{Brinsfield2009} &  \\
10772 & 1990 YM & 2006 3 -- 2006 3 & Koff & Antelope Hills Observatory, Bennett (H09) \\
      &         & 2006 4 -- 2006 4 & Warner & Palmer Divide Observatory (716) \\
31383 & 1998 XJ$_{94}$ & 2006 4 -- 2006 4 & \citet{Higgins2006} &  \\
\end{longtable}
\tablefoot{
\tablefoottext{a}{On line at \texttt{http://www.david-higgins.com/Astronomy/asteroid/lightcurves.htm}} 
\tablefoottext{b}{On line at \texttt{http://aslc-nm.org/Pilcher.html}} 
\tablefoottext{c}{Observations, reductions, and calibration methods are described in \citet{Polishook2008, Polishook2009}} 
}
}
}

\end{document}